\documentclass[12pt]{article}
\pdfoutput=1
\usepackage{jheppub}

\usepackage{amsmath,amssymb,amsfonts}
\usepackage{graphicx}
\usepackage{tikz}
\usepackage{dsfont}
\usepackage{comment}

\usepackage{booktabs}
\usepackage{multirow}

\def\bZ{\mathbb{Z}}

\usepackage{subfigure}

\usepackage{tikz}

\makeatletter\def\l@subsubsection#1#2{}%
\makeatother

\def\be{\begin{eqnarray}}
\def\ee{\end{eqnarray}}

\begin{document}

\setcounter{table}{0}

\preprint{IPMU20-0016}

\title{Generalized symmetries and holography in ABJM-type theories}

\author[a]{Oren Bergman,}
\emailAdd{bergman@physics.technion.ac.il}
\author[b]{Yuji Tachikawa,}
\emailAdd{yuji.tachikawa@ipmu.jp}
\author[b,c]{Gabi Zafrir} 
\emailAdd{gabi.zafrir@unimib.it}

\affiliation[a]{Department of Physics, Technion, Israel Institute of Technology\\
Haifa, 32000, Israel\\[-4mm]}

\affiliation[b]{Kavli Institute for the Physics and Mathematics of the Universe (WPI), University of Tokyo\\
Kashiwa, Chiba 277-8583, Japan\\[-4mm]
}

\affiliation[c]{Dipartimento di Fisica, Universit\`a di Milano-Bicocca \& INFN, \\ 
Sezione di Milano-Bicocca, I-20126 Milano, Italy}

\abstract{We revisit the ${\cal N}=6$ superconformal Chern-Simons-matter theories and their supergravity duals in the context of generalized symmetries.
This allows us to finally clarify how the $SU(N)\times SU(N)$ and $(SU(N)\times SU(N))/\mathbb{Z}_N$ theories,
as well as other quotient theories that have recently been discussed, fit into the holographic framework.
It also resolves a long standing puzzle regarding the di-baryon operator in the $U(N)\times U(N)$ theory.}

\maketitle

\section{Introduction}

Extended operators and the generalized symmetries acting on them 
provide a new viewpoint on various structures in quantum field theory \cite{Aharony:2013hda,Gaiotto:2014kfa}. 
In many cases they are the natural language to describe these structures, and in some instances they provide genuinely new insights.
For example this has led to a more detailed understanding of duality in 4d ${\cal N}=4$ SYM theory,
where discrete one-form symmetries and the line operators they act on reveal an intricate structure of duality orbits \cite{Aharony:2013hda}.
Furthermore, through the process of gauging the one-form symmetries or their subgroups one can relate all ${\cal N}=4$ theories with a given 
gauge algebra \cite{Gaiotto:2014kfa}.

Higher form symmetries also appear naturally in string theory and are therefore relevant in holographic descriptions of quantum field theories.
In particular the one-form symmetries of 4d ${\cal N}=4$ SYM theories with gauge algebra $su(N)$ descend from the two-form symmetries of Type IIB 
string theory on $AdS_5\times S^5$ associated to the two-form gauge fields $B$ and $C$.
From the 5d bulk point of view the different 4d theories correspond to different boundary conditions imposed on these two fields 
at the boundary of $AdS_5$,
and the $SL(2,\mathbb{Z})$ duality action in the field theory corresponds to the $SL(2,\mathbb{Z})$ duality action in Type IIB string theory on the doublet $(B,C)$ \cite{Witten:1998wy}.
A crucial observation of  \cite{Witten:1998wy} was that the allowed boundary conditions on $B$ and $C$ are constrained by 
a topological term in the 5d low-energy effective theory,
\be
\label{IIBAction}
S_{top} = \frac{N}{2\pi} \int_{X_5} B\wedge dC \,,
\ee
which is the dominant term near the boundary of $AdS_5$.\footnote{See also \cite{Hofman:2017vwr} for a more recent discussion.}
By itself this action describes a theory with a $\mathbb{Z}_N$ one-form gauge symmetry.
The field strengths of $B$ and $C$ are trivial, but the potentials may have non-trivial holonomies taking values in $\mathbb{Z}_N$.
In the quantum theory the holonomies of $B$ and $C$ are canonically conjugate variables spanning a discrete phase space.
The simplest boundary conditions correspond to fixing $B$ at the boundary while allowing the boundary value of $C$ to be free
as an element of $\mathbb{Z}_N$, or to fixing $C$ at the boundary while allowing the boundary value of $B$ to be free
as an element of $\mathbb{Z}_N$. These two sets of boundary conditions 
correspond respectively to the 4d field theories with gauge groups $SU(N)$ and $SU(N)/\mathbb{Z}_N$, 
which as we know are related by S-duality.
More general boundary conditions, and their relation to the 4d field theories were discussed in \cite{Gaiotto:2014kfa}.

In many respects, the three-dimensional version of 4d ${\cal N}=4$ SYM theory is 3d ${\cal N}=6$ Super-Chern-Simons (SCS) theory,
a class of 3d superconformal theories with twelve Poincar\'e supersymmetries.
Originally three such theories were identified in \cite{Aharony:2008ug}.
The three are based on the gauge groups and Chern-Simons (CS) levels given by $U(N)_k\times U(N)_{-k}$, $SU(N)_k\times SU(N)_{-k}$,
and $(SU(N)_k\times SU(N)_{-k})/\mathbb{Z}_N$, all 
containing matter fields in the $({\bf N},\bar{\bf N})$
representation corresponding in the language of 3d ${\cal N}=4$ supersymmetry to two hypermultiplets.
The latter two theories are generalizations of the BLG theories, which correspond to the $N=2$ cases 
\cite{Bagger:2006sk,Gustavsson:2007vu}.\footnote{These theories were originally formulated in terms of a Lie 3-algebra, but were subsequently 
shown to be equivalent to CS gauge theories in \cite{VanRaamsdonk:2008ft}.}
The $U\times U$ theory was singled out, via its embedding in string theory, as the theory describing $N$ M2-branes in 
the particularly simple geometrical background given by $\mathbb{R}^{1,2}\times \mathbb{C}^4/\mathbb{Z}_k$.
This also implied that at large $N$ the $U\times U$ theory is dual to M theory on $AdS_4\times S^7/\mathbb{Z}_k$,
or equivalently to Type IIA string theory on $AdS_4\times \mathbb{C}P^3$, providing the first explicit realization of
AdS/CFT in these dimensions.
On the other hand the holographic duals of the $SU\times SU$ theories have remained mysterious, since 
these theories do not appear in general to describe M2-branes in an eleven-dimensional geometry.\footnote{For $N=2$ and $k=1$
the two $SU\times SU$ theories turn out to be equivalent to the $U(2)_2\times U(2)_{-2}$ and $U(2)_1\times U(2)_{-1}$ theory,
respectively \cite{Lambert:2010ji,Bashkirov:2011pt}, and therefore describe two M2-branes in $\mathbb{R}^8/\mathbb{Z}_2$
and $\mathbb{R}^8$, respectively. A generalization of the second equivalence, conjectured in \cite{Lambert:2010ji}, will be discussed below.}

All 3d ${\cal N}=6$ SCS theories were subsequently classified, up to discrete quotients, in \cite{Schnabl:2008wj}.
A more complete classification has recently appeared in \cite{Tachikawa:2019dvq}.
This includes theories with gauge groups $(U(N)_k\times U(N)_{-k})/\mathbb{Z}_{m'}$, where $m'$ is a divisor of $k$, and 
$(SU(N)_k\times SU(N)_{-k})/\mathbb{Z}_{n'}$, where $n'$ is a divisor of $N$,
which interpolate in some sense between the $SU\times SU$ theory and the $U\times U$ theory.
The authors of  \cite{Tachikawa:2019dvq} have also shown that the first theory with $m'=k$ is non-perturbatively equivalent to
the second theory with $n'=N$, verifying the conjecture of \cite{Lambert:2010ji}.
The situation now resembles that of the 4d ${\cal N}=4$ theories with $su(N)$ algebra.
The different 3d ${\cal N}=6$ theories are related by gauging either a discrete one-form symmetry or a discrete zero-form symmetry.

The main question we wish to address in this paper is how the different 3d ${\cal N}=6$ theories fit into the dual supergravity picture.
The role played by the $U(N)_k\times U(N)_{-k}$ theory  
has remained special in that it is the only one that has a geometric interpretation in terms of M2-branes.
All other theories do not appear to have a clear interpretation in terms of M2-branes,
as their moduli spaces involve non-geometric quotients.
Nevertheless, as we will show, they do have a simple large $N$ holographic dual.
Similar to the case of the 4d ${\cal N}=4$ theories, the different 3d ${\cal N}=6$ theories will correspond to different boundary 
conditions imposed on a set of gauge fields in $AdS_4$.
The main ingredient will again be a topological term in the supergravity action, this time given by
\be
S_{top} = \frac{1}{2\pi} \int_{X_4} B\wedge d(NA_{D0} + kA_{D4}) \,,
\label{bulktop}
\ee
where, in the Type IIA string theory description, $B$ is the NSNS two-form gauge field, 
and $A_{D0}$ and $A_{D4}$ are one-form gauge fields originating in the RR sector.
The different boundary conditions for $(B,A_{D0},A_{D4})$ allowed by this term correspond to different 3d ${\cal N}=6$ theories.
We will show that a subset of these is given by the ${\cal N}=6$ SCS theories listed above.
In particular we will identify the boundary conditions in $AdS_4$ corresponding to the $U(N)_k\times U(N)_{-k}$ theory,
something that was not explicitly done in \cite{Aharony:2008ug}.

One of the original motivations for this work has been a long-standing puzzle about the existence of di-baryon states in the supergravity dual
of the ${\cal N}=6$ SCS theory 
\cite{Aharony:2008ug}. Namely there exists a state in $AdS_4$ corresponding to a wrapped D4-brane
that has the properties of a di-baryon operator in the 3d field theory, even though in the $U(N)_k\times U(N)_{-k}$ theory
it is not a gauge invariant operator. 
As we will see, the identification of the correct boundary conditions leads to a simple resolution of the puzzle.\footnote{The di-baryon question
was previously addressed in \cite{Park:2008bk} and in \cite{Berenstein:2009sa}. 
Our resolution is different.
We also note that the holography of $2N$ D3-branes on an O3$^-$-plane leads to a Pfaffian puzzle.
Namely, a D3-brane wrapped on $\mathbb{R}P^3\subset \mathbb{R}P^5$ has the correct property to be identified with the Pfaffian operator of the boundary 4d $\mathcal{N}=4$ $SO(2N)$ theory,
whose moduli space does not admit a simple interpretation in terms of $2N$ D3-branes moving on this background.
Rather, such a brane interpretation requires the gauge group to be $O(2N)$, for which no Pfaffian operator exists.
This point was resolved in \cite[Sec.~3.3]{Aharony:2016kai} using the boundary condition of the bulk discrete gauge field.}

The rest of the paper is organized as follows.
In section 2 we will discuss 3d ${\cal N}=6$ SCS theories, highlighting their generalized global symmetries and spectrum of local and line operators.
In section 3 we will describe their supergravity duals and determine the correspondence between a subset of the allowed boundary conditions and 
the 3d ${\cal N}=6$ SCS theories.
Section 4 contains our conclusions and a number of open questions for the future.
There is also an appendix reviewing monopole operators in 3d gauge theories.

\section{${\cal N}=6$ Super-Chern-Simons Theories}

We will first concentrate on the four ``basic'' 3d ${\cal N}=6$ theories, that are in some sense the analogs of the the 4d ${\cal N}=4$ $SU(N)$ and
$SU(N)/\mathbb{Z}_N$ theories. 
The 3d theories are based on the gauge groups and CS levels given by 
$U(N)_k\times U(N)_{-k}$, $SU(N)_k\times SU(N)_{-k}$, $(U(N)_k\times U(N)_{-k})/\mathbb{Z}_k$, and $(SU(N)_k\times SU(N)_{-k})/\mathbb{Z}_N$,
and all have bi-fundamental matter fields corresponding to two hypermultiplets.
Then we will discuss the more general set of theories that were found in \cite{Tachikawa:2019dvq}.


\subsection{$U(N)_k\times U(N)_{-k}$}

This is the theory originally featured in \cite{Aharony:2008ug}.

As is well known by now, a 3d $U(N)$ gauge theory has a $U(1)$ global symmetry generated by the topological current $j=*\mbox{Tr}f$.
In our case there are two such currents $j_1, j_2$ corresponding to the two $U(N)$ gauge fields $a_1, a_2$.
We denote the corresponding charges by $m_1, m_2$.
This symmetry acts only on monopole operators, which are defined as local operators that insert
a magnetic flux on the 2-sphere that surrounds them (see the appendix for a brief review of monopole operators). 

For a generic monopole the magnetic fluxes are given by
\be
h_i = \int_{S^2} \frac{f_i}{2\pi} = \mbox{diag}(m^1_i,\ldots, m^N_i)\,,
\ee
where $i=1,2$ labels the two $U(N)$ factors and $m^a_i$, with $a=1,\ldots , N$, 
are integers. The Weyl transformations allow us to choose $m^1_i\geq \cdots \geq m^N_i$. 
This operator carries charges under the two topological $U(1)$ symmetries given by $m_1 = \sum_a m^a_1$
and $m_2 = \sum_a m^a_2$.
A monopole operator inserting these fluxes will be denoted succinctly by $T_{m_1,m_2}$.

Due to the CS terms the monopoles also carry non-trivial gauge charges.
In particular the basic monopole $T_{1,0}$
transforms in the $({\bf N})^k_{sym}$ representation
of the first $U(N)$ factor (and therefore carries $k$ units of charge under the $U(1)$ part), and the fundamental monopole $T_{0,1}$
transforms in the $(\bar{\bf N})^k_{sym}$ representation of the second $U(N)$ factor (and therefore carries $-k$ units of charge under the $U(1)$ part).
So generically these are not gauge invariant operators, and are not part of the physical spectrum.
However for a special class of monopole operators we can form gauge invariant operators by ``dressing" the monopoles with 
the matter fields. Since the latter transform in the $({\bf N},\bar{\bf N})$ representation, this is only possible for monopole operators with $m_1=m_2$.
Specifically the BPS operators are built from monopoles defined by $m_1^1=m_2^1=m$ and 
$m_i^{a>1}=0$, and are given by 
\be 
\label{dressedmonopole}
{\cal M}_m = T_{m,m} \cdot (\phi^\dagger)^{mk}_{sym} \,,
\ee 
where $\phi$ denotes the four complex scalar components of the matter multiplet.
Monopole operators with $m_1\neq m_2$ cannot be dressed into gauge invariant operators.
For example the operator $T_{1,-1}$ (which in our convention means $m_1^1=-m_2^N=1$) 
transforms as $({\bf N})^k_{sym}$ under both $U(N)$ factors, and so cannot be dressed into a gauge invariant operator.
Therefore only the symmetric combination of the two topological $U(1)$ symmetries
generated by $j_1 + j_2$ acts nontrivially on the physical spectrum.

The complete spectrum of BPS operators is given by 
(\ref{dressedmonopole}) combined with neutral mesonic operators of the form $\mbox{Tr}(\phi^\dagger\phi)^n$.
One can also form a di-baryon operator using an antisymmetric product of bi-fundamental fields
\be
\label{dibaryon}
{\cal B} = \mbox{det}\, \phi = \epsilon_{a_1\cdots a_N} \epsilon^{b_1\cdots b_N} \phi^{a_1}_{b_1} \cdots \phi^{a_N}_{b_N} \,.
\ee
This is invariant under $SU(N)\times SU(N)$ but carries charges $(N,-N)$ under the $U(1)$ factors.
It seems one could obtain a gauge invariant operator by dressing the di-baryon with a monopole defined by
\be
\label{1/kMonopole}
h_1 = h_2 = \mbox{} - \mbox{diag}\left(\frac{1}{k},\ldots,\frac{1}{k}\right) \,.
\ee
However this violates the Dirac quantization condition, and is therefore forbidden.
The $k$-fold product of di-baryons can be properly dressed with integer fluxes, but the resulting operator is equivalent to an $N$-fold product of dressed monopoles,
\be 
\label{rearrangement}
(\mbox{det}\,\phi)^k \cdot T_{\{-1,\dots, -1;-1, \ldots, -1\}} \sim [T_{-1,-1} \cdot (\phi^{k})_{sym}]^N \,,
\ee
and so does not represent an independent gauge invariant local operator.

The $U(N)_k\times U(N)_{-k}$ theory has two other interesting properties that were not discussed in \cite{Aharony:2008ug},
and which will play an important role in what follows.

\paragraph{One-form symmetry:}
The $U(N)_k\times U(N)_{-k}$ theory has a global $\mathbb{Z}_{k}$ one-form symmetry
acting on a subset of Wilson line operators.
To see this, let us first consider a single $U(N)_{k} = (SU(N)_k\times U(1)_{Nk})/\mathbb{Z}_N$.
The $SU(N)_k$ theory has a $\mathbb{Z}_N$ one-form symmetry, and the $U(1)_{Nk}$ theory has a $\mathbb{Z}_{Nk}$ one-form symmetry.
Modding out a combined $\mathbb{Z}_N$ then leaves just $\mathbb{Z}_k$ for a single $U(N)_k$.
The Wilson line in the representation $\mathbf{N}$ has charge 1, 
and a collection of $k$ such Wilson lines can be screened by a unit monopole operator $T_1$. 

In our situation we have $U(N)_k$ and $U(N)_{-k}$,
which naively give us a $\mathbb{Z}_k\times \mathbb{Z}_k$ one-form symmetry.
However the anti-diagonal combination is absent due to the presence of matter fields in the $(\mathbf{N},\bar{\mathbf{N}})$ representation:
a Wilson line in the representation $(\mathbf{N},\bar{\mathbf{N}})$ can be screened by a single matter field operator.
On the other hand a Wilson line in the representation $(\mathbf{N},\mathbf{N})$ cannot be screened.
This line carries one unit of charge under the diagonal $\mathbb{Z}_k$, and 
a collection of $k$ such lines can be screened by the monopole operator $T_{1,-1}$.

\paragraph{A mixed anomaly:}
The background field for the $\mathbb{Z}_k$ one-form symmetry is a degree-2 $\mathbb{Z}_k$-valued cohomology class $\mathsf{B}$.
In particular, the configuration $\int_{S^2} \mathsf{B} = j\in \mathbb{Z}_k$ 
is equivalent to a monopole with the fluxes
\be
h_1 = h_2 = j\, \mbox{diag}\left(\frac1k,\ldots, \frac1k\right)\,.
\label{Zk-background}
\ee
This implies that the $U(1)$ zero-form symmetry and the $\mathbb{Z}_k$ one-form symmetry have a mixed anomaly, 
for the rather trivial reason that the above monopole 
in general carries a fractional $U(1)$ charge $m=jN/k$.
In the presence of a nontrivial background $U(1)$ field $A$,
the partition function then has a phase ambiguity of $e^{2\pi ij N/k}$,
signaling the presence of the anomaly.
The 4d characteristic class which describes this anomaly is 
\begin{equation}
\exp\left(2\pi i \frac{N}{k}\int \mathsf{B}\, c_1(F) \right)\,,
\label{Zk-anomaly}
\end{equation} 
where $F=dA$.

\subsection{$SU(N)_k\times SU(N)_{-k}$}

This theory differs from the $U(N)_k\times U(N)_{-k}$ theory in that the $U(1)$'s are not gauged.
The anti-diagonal $U(1)$ acts non-trivially on the matter fields and therefore defines a baryonic $U(1)$ zero-form symmetry in the theory.
Let us be more precise about the periodicity of this $U(1)$ symmetry group.
The basic gauge invariant operator charged under the baryonic $U(1)$ is the di-baryon ${\cal B}$ \eqref{dibaryon}.
If we assign charge 1 to $\phi$, $\cal B$ has charge $N$.
But we can in fact assign charge 1 to ${\cal B}$ in the following sense.
Pick $g\in U(1)$, and say that under this element we have the transformation
\begin{equation}
\mathcal{B} \mapsto g\mathcal{B}. \label{gB}
\end{equation}
For this operation we need to assign the transformation rule $\phi \mapsto g^{1/N} \phi$.
This action has an ambiguity by $N$-th roots of unity.
However, a multiplication by $N$-th roots of unity is part of the $SU(N)$ gauge symmetry,
and therefore the transformation \eqref{gB} is well-defined
at the level of the elementary fields in the Lagrangian.

There are no topological $U(1)$ symmetries in this theory, and correspondingly the monopole operators (see appendix) 
do not carry a conserved charge.

\paragraph{One-form symmetry:}
There is a $\mathbb{Z}_N$ one-form symmetry acting on a subset of Wilson line operators, where the basic one is again in
the representation $({\bf N},{\bf N})$.
As in four dimensions $N$ such Wilson lines can be screened by a gluon.
There is not a second $\mathbb{Z}_N$ one-form symmetry acting on the $({\bf N},\bar{\bf N})$ Wilson line,
since that is again screened by the bi-fundamental field.

\paragraph{A mixed anomaly:}
The $U(1)$ zero-form symmetry and the $\mathbb{Z}_N$ one-form symmetry have a mixed anomaly due to a mechanism similar to the one we saw 
in the $U\times U$ theory above.
This time, the background field is a degree-2 $\mathbb{Z}_N$-valued cohomology class $\mathsf{B}$,
and the configuration $\int_{S^2} \mathsf{B} = j \in \mathbb{Z}_N$ 
is equivalent to a monopole with fluxes
\be
h_1=h_2 = j\, \mbox{diag}\left(1-\frac1N,-\frac1N\ldots, -\frac1N \right).
\label{ZN-background}
\ee
Due to the Chern-Simons term, 
it has the gauge charge $((\mathbf{N})^k_{sym},(\bar{\mathbf{N}})^k_{sym})$,
which can be made gauge invariant by attaching a symmetric product of $k$ bi-fundamental fields.
This however has a fractional baryonic $U(1)$ charge $k/N$ in our normalization where $\mathcal{B}$ has charge 1.
In the presence of background fields for both the one-form and zero-form symmetries, $\mathsf{B}$ and $A$,
this introduces a phase ambiguity $e^{2\pi i j k/N}$ in the partition function.
The 4d characteristic class which describes this anomaly is \begin{equation}
\exp\left(2\pi i \frac{k}{N}\int \mathsf{B} \,c_1(F) \right).
\label{ZN-anomaly}
\end{equation}


\subsection{$(U(N)_k\times U(N)_{-k})/\mathbb{Z}_k$}

This theory corresponds to gauging the $\mathbb{Z}_k$ one-form symmetry of the $U(N)_k\times U(N)_{-k}$ theory.
This removes Wilson lines in representations with non-trivial diagonal $k$-ality.
In other words only Wilson lines in the representations $({\bf N}^{k},{\bf N}^{k})$ and 
$({\bf N},\bar{\bf N})$, and their products, are kept. The latter, as we recall, are screened by the matter fields, and the former by the monopole operator
of the form $T_{1,-1}$.
There are no unscreened Wilson lines, and correspondingly there is no remaining one-form symmetry.
At the same time, the theory admits additional monopole operators corresponding to fractional magnetic fluxes of the form \eqref{1/kMonopole},
that in turn allow us to dress the di-baryon operator into a gauge invariant operator,
\be
{\cal B} := T_{\{-\frac{1}{k},\ldots,-\frac{1}{k};-\frac{1}{k},\ldots, -\frac{1}{k}\}} \cdot \mbox{det}\,\phi .
\ee
The relation (\ref{rearrangement}) still holds, and implies here a chiral-ring like relation involving the gauge invariant dressed di-baryons
and dressed monopole operators,
\be
\label{ChiralRing}
{\cal M}_{-N} = {\cal B}^k   \,.
\ee

This theory turns out to have a rather intricate global symmetry structure, that we shall next explore. First, on physical grounds, we expect two zero-form symmetries. 
One is just the topological $U(1)$ symmetry that exists prior to the gauging of the $\mathbb{Z}_k$ one-form symmetry. 
However, after the gauging we expect to gain a new $\mathbb{Z}_k$ zero-form symmetry which acts on the newly added monopole operators. 
In particular, it should also act on the dressed di-baryon ${\cal B}$. 
At this point we might be tempted to say that the global zero-form symmetry is $U(1) \times \mathbb{Z}_k$, but that turns out to be not quite right. 
The issue is that there is the possibility that the $\mathbb{Z}_k$ is not independent, but rather part of the $U(1)$. 
Specifically, we seek a $U(1)$ transformation that acts trivially on ${\cal M}_{1}$, but acts on ${\cal B}$ like a $\mathbb{Z}_k$ zero-form symmetry. 
If such a transformation exists then the $\mathbb{Z}_k$ is actually contained in the $U(1)$. 
Under the $U(1)$, ${\cal M}_{1}$ has charge $1$ while ${\cal B}$ has charge $-\frac{N}{k}$.
As a result,  the action of  elements in $\mathbb{Z}_k$ except for its $\mathbb{Z}_{\gcd(N,k)}$ subgroup can be reproduced using the $U(1)$ action.
Thus, we conclude that the zero-form global symmetry is $U(1) \times \mathbb{Z}_{\gcd(N,k)}$.

It will be beneficial for us later to consider the structure of the global symmetry from a different viewpoint. For that we temporarily introduce two $U(1)$ symmetries,
$U(1)_{\cal M} \times U(1)_{\cal B}$,
under which ${\cal M}_{m}$ has charge $(m,0)$ and ${\cal B}^\ell$ has charge $(0,\ell)$.
We denote the group element by $(g_{\cal M},g_{\cal B})$.
The chiral-ring-like relation \eqref{ChiralRing} imposes the constraint that $
g_{\cal M}^N g_{\cal B}^k =1.
$
Therefore the zero-form symmetry $G_0$ of this theory is the subgroup of $U(1)^2$ specified as follows \begin{equation}
G_0 := \{ (g_{\cal M},g_{\cal B} ) \mid g_{\cal M}^N g_{\cal B}^k =1 \} \subset U(1)_{\cal M}\times U(1)_{\cal B}. \label{G0}
\end{equation}

This constraint reduces the continuous part to a single $U(1)$ which can be chosen to be the previously defined one. However, additionally we also have the discrete transformations, $\mathbb{Z}_N \subset U(1)_{\cal M}$ and $\mathbb{Z}_k \subset U(1)_{\cal B}$, but from these we need to mod out the part that is included in the $U(1)$. By the previous argument this leaves us with only a $\mathbb{Z}_{\gcd(N,k)}$ discrete symmetry. Note that in the previous argument we naturally chose to present $G_0$ as $U(1) \times \mathbb{Z}_{\gcd(N,k)}$ with $\mathbb{Z}_{\gcd(N,k)} \subset \mathbb{Z}_k \subset U(1)_{\cal B}$.
But it should be apparent that we could also present $G_0$ as $U(1) \times \mathbb{Z}_{\gcd(N,k)}$ with $\mathbb{Z}_{\gcd(N,k)} \subset \mathbb{Z}_N \subset U(1)_{\cal M}$ and the $U(1)$ now defined so that it acts on ${\cal B}$ with charge $1$ and on ${\cal M}_{1}$ with charge $-\frac{k}{N}$. Thus, while $G_0$ is $U(1)\times \mathbb{Z}_{\gcd(N,k)}$ as a group, there is no canonical way to choose the $\mathbb{Z}_{\gcd(N,k)}$ part.

We can describe this more formally as follows. First, we introduce the integers $p,q$ by \begin{equation}
N=p\gcd(N,k),\qquad k=q \gcd(N,k).
\end{equation} 
We have a natural embedding $U(1)\to G_0$ given by \begin{equation}
U(1)\ni g \mapsto (g^{-q},g^p)\in G_0
\end{equation} and the natural projection $G_0\to \mathbb{Z}_{\gcd(N,k)}$ given by \begin{equation}
G_0\ni (g_{\cal M},g_{\cal B}) \mapsto g_{\cal M}^p g_{\cal B}^q \in \mathbb{Z}_{\gcd(N,k)}.
\end{equation}
These two operations make $G_0$ a group extension \begin{equation}
0\to U(1)\to G_0 \to \mathbb{Z}_{\gcd(N,k)}\to 0.\label{gcdNk-ext}
\end{equation}
We can split $G_0$ as  $G_0\simeq U(1)\times \mathbb{Z}_{\gcd(N,k)}$ 
but there are multiple ways to do this. 

We can get back to the $U(N)_k\times U(N)_{-k}$ theory by gauging 
the $\mathbb{Z}_k$ subgroup 
of $G_0$ generated by the element $(g_{\cal M},g_{\cal B})=(1,e^{2\pi i/k})$.
This removes the di-baryon; 
the remaining zero-form symmetry is $G_0/\mathbb{Z}_k\simeq U(1)_\mathcal{M}$;
here we are utilizing the extension \begin{equation}
0\to \mathbb{Z}_k \to G_0 \to U(1)_{\cal M} \to 0 \label{Zk-ext} 
\end{equation} instead of \eqref{gcdNk-ext}.
Now, the gauging introduces a $\mathbb{Z}_k$ gauge field, and therefore a global $\mathbb{Z}_k$ one-form symmetry.
It is a general fact \cite{Tachikawa:2017gyf} that the gauging of a finite subgroup of an extension such as \eqref{Zk-ext}
results in a mixed anomaly \eqref{Zk-anomaly}.

\subsection{$(SU(N)_k\times SU(N)_{-k})/\mathbb{Z}_N$}

This theory is obtained by gauging the $\mathbb{Z}_N$ one-form symmetry 
of the $SU(N)_k\times SU(N)_{-k}$ theory.
This removes Wilson lines in representations with non-trivial diagonal $N$-ality.
In other words only Wilson lines in the representations $({\bf N}^{N},{\bf N}^{N})$ and 
$({\bf N},\bar{\bf N})$, and their products, are kept. The latter, as we recall, are screened by the matter fields, and the former by the gluons.
So, as above, there are no unscreened Wilson lines and no remaining one-form symmetry.
At the same time the theory admits monopole operators with fractional magnetic fluxes of the form (\ref{ZN-background}).
These can be dressed into gauge-invariant operators with matter fields as follows
\be 
\label{SUSUdressedmonopole}
{\cal M}_j = T_{j\{1-\frac{1}{N},-\frac{1}{N},\dots,-\frac{1}{N};1-\frac{1}{N},-\frac{1}{N},\dots,-\frac{1}{N}\}} \cdot (\phi^\dagger)^{jk}_{sym} \,.
\ee 
The chiral-ring like relation (\ref{ChiralRing}) holds in this theory as well. 

The analysis of the global symmetry of this theory turns out to be rather similar to the previous case. 
In this case we expect a $\mathbb{Z}_N$ zero-form symmetry acting on the newly added monopoles, in addition to the $U(1)$ acting on the di-baryons.
However, like in the previous case, these symmetries are not independent. Specifically, the basic gauge invariant operators include the basic di-baryon
${\cal B}$, which is neutral under $\mathbb{Z}_N$ and has charge $1$ under the $U(1)$, and the basic dressed monopole ${\cal M}_1$ which is acted on by the generator of $\mathbb{Z}_N$, and has charge $-\frac{k}{N}$ under the $U(1)$. We then again see that only the $\mathbb{Z}_{\gcd(N,k)}$ part of $\mathbb{Z}_N$ is independent. Therefore, we again find a $U(1)\times \mathbb{Z}_{\gcd(N,k)}$ global symmetry. In fact this theory has the same global symmetry structure, $G_0$, as the previous case, only that here we have naturally chosen a different decomposition as a $U(1)\times \mathbb{Z}_{\gcd(N,k)}$ group,
with $\mathbb{Z}_{\gcd(N,k)} \subset \mathbb{Z}_N \subset U(1)_{\cal M}$.


We can get back to the $SU(N)_k\times SU(N)_{-k}$ theory by gauging 
the $\mathbb{Z}_N$ subgroup 
of $G_0$ generated by the element $(g_{\cal M},g_{\cal B})=(e^{2\pi i/N},1)$,
which forms the extension \begin{equation}
0\to \mathbb{Z}_N \to G_0 \to U(1)_{\cal B} \to 0. \label{ZN-ext} 
\end{equation}
This removes the dressed monopoles, reducing the zero-form global symmetry from $G_0$ to $U(1)_{\cal B}$, and
at the same time introduces a $\mathbb{Z}_N$ gauge field, and therefore a global $\mathbb{Z}_N$ one-form symmetry.
Again the general argument of \cite{Tachikawa:2017gyf} implies that there is the mixed anomaly \eqref{ZN-anomaly}.

The fact that the global symmetry and the spectrum of the $(SU(N)_k\times SU(N)_{-k})/\mathbb{Z}_N$ theory  is identical to that of the $(U(N)_k\times U(N)_{-k})/\mathbb{Z}_k$
 theory is not accidental.
These two theories are in fact equivalent, as was shown in \cite{Tachikawa:2019dvq};
one simply needs to integrate out the $u(1)\times u(1)$ part with care.

\subsection{Generalization}

\begin{figure}[h!]
\center
\includegraphics[height=0.15\textwidth]{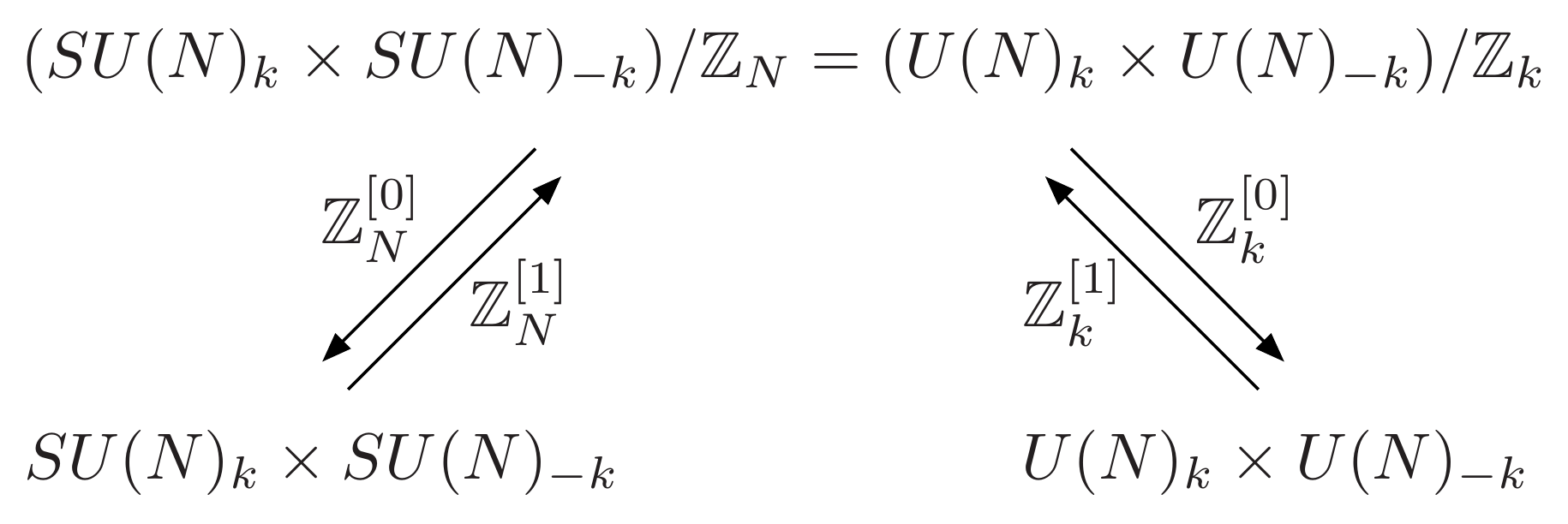} 
\caption{The discrete gauging relations between the three basic 3d ${\cal N}=6$ theories.}
\label{DiscreteGauging1}
\end{figure}

The four, or really three, basic theories we discussed above are related via gauging a discrete symmetry,
which is either $\mathbb{Z}_k$ or $\mathbb{Z}_N$, and either a zero-form symmetry or a one-form symmetry, Fig.~\ref{DiscreteGauging1}.
We can generalize this procedure by gauging a subgroup of the relevant discrete symmetry.
This produces the set of ${\cal N}=6$ theories found in \cite{Tachikawa:2019dvq}.

A good starting point is the $(SU(N)_k\times SU(N)_{-k})/\mathbb{Z}_N$, or
$(U(N)_k\times U(N)_{-k})/\mathbb{Z}_k$, theory.
This theory has a global zero-form symmetry $G_0=U(1)\times \mathbb{Z}_{\gcd(N,k)}$.
There are two simple ways to proceed from here.

If $k=mm'$ we can gauge $\mathbb{Z}_{m}\subset \mathbb{Z}_k \subset G_0$. 
The resulting theory has a gauge group $(U(N)_k\times U(N)_{-k})/\mathbb{Z}_{m'}$.
From the point of view of the $(U(N)_k\times U(N)_{-k})/\mathbb{Z}_k$ description of the original theory, 
the discrete gauging removes the additional monopole operators whose fluxes are not multiples of $\frac{m}{k}$,
and at the same time introduces Wilson line operators in the representation $({\bf N}^{m'},{\bf N}^{m'})$, and its multiples.
These are naturally charged under the resulting $\mathbb{Z}_m$ one-form symmetry, since $m$ copies of the basic Wilson line
can be screened by the monopole operator $T_{1,-1}$.
The zero-form symmetry of this theory is the commutant of $\mathbb{Z}_m$ in $G_0=U(1)\times \mathbb{Z}_{\gcd(N,k)}$,
which is $U(1)\times  \mathbb{Z}_{\gcd(N,m')}$.
The dressed monopole operators, which are unaffected by the discrete gauging, carry integer charges under $U(1)$
and are neutral under $\mathbb{Z}_{\gcd(N,m')}$.
The di-baryon operators now come in multiples of $m$, carry a $U(1)$ charge that is an integer multiple of $mN/k$,
and are charged under $\mathbb{Z}_{\gcd(N,m')}$.
Alternatively, we can also define this theory by starting with the $U(N)_k\times U(N)_{-k}$ theory and gauging
a $\mathbb{Z}_{m'}$ subgroup of the $\mathbb{Z}_k$ one-form symmetry (see Fig.~\ref{DiscreteGauging2}).
This removes from the spectrum Wilson lines in representations that are not $m'$-multiples of $({\bf N},{\bf N})$,
leaving a $\mathbb{Z}_m$ one-form symmetry acting on the remaining Wilson lines.
At the same time this introduces additional monopole operators with magnetic fluxes given by integer multiples of $1/m'=m/k$,
which can in turn be used to dress $m$ multiples of the di-baryon operator.
The dressed-monopole and di-baryon operators satisfy the relation (\ref{ChiralRing}), now usefully expressed as
\be
{\cal M}_{-N} = {\cal B}^{mm'} \,.
\ee

If $N=nn'$ we can gauge a $\mathbb{Z}_{n}\subset \mathbb{Z}_N \subset G_0$. 
The resulting theory has a gauge group $(SU(N)_k\times SU(N)_{-k})/\mathbb{Z}_{n'}$.
From the point of view of the $(SU(N)_k\times SU(N)_{-k})/\mathbb{Z}_N$ description of the original theory, 
the discrete gauging removes dressed monopole operators of the form (\ref{SUSUdressedmonopole}) with $j$ not a multiple of $n$,
and at the same time introduces Wilson lines in the representation $({\bf N}^{n'},{\bf N}^{n'})$, and its multiples.
These are naturally charged under the resulting $\mathbb{Z}_n$ one-form symmetry, since $n$ copies of the basic Wilson line
can be screened by the gluons.
The zero-form symmetry of this theory is the commutant of $\mathbb{Z}_n$ in $G_0=U(1)\times \mathbb{Z}_{\gcd(N,k)}$,
which is $U(1)\times  \mathbb{Z}_{\gcd(n',k)}$.
The di-baryon operators, which are unaffected by the discrete gauging, carry an integer charge under $U(1)$
and are neutral under $\mathbb{Z}_{\gcd(n',k)}$.
The dressed monopole operators now come in multiples of $n$, carry a $U(1)$ charge that is an integer multiple of $nk/N$,
and are charged under $\mathbb{Z}_{\gcd(n',k)}$.
Alternatively, we can also define this theory by starting with the $SU(N)_k\times SU(N)_{-k}$ theory and gauging
a $\mathbb{Z}_{n'}$ subgroup of the $\mathbb{Z}_N$ one-form symmetry (see Fig.~\ref{DiscreteGauging2}).
This removes Wilson lines in representations that are not $n'$-multiples of $({\bf N},{\bf N})$,
leaving a $\mathbb{Z}_n$ one-form symmetry acting on the remaining Wilson lines.
At the same time it introduces monopole operators of the form (\ref{SUSUdressedmonopole})
with $j$ a multiple of $n$.
The dressed-monopole and di-baryon operators satisfy the relation (\ref{ChiralRing}), now usefully expressed as
\be
{\cal M}_{-nn'} = {\cal B}^k \,.
\ee

\tikzset{>=latex}
\begin{figure}[h!]
\center \scriptsize
\begin{tikzpicture}
\node[anchor=west] (A) at (0,0) {$(U(N)_k\times U(N)_{-k})/\mathbb{Z}_k$};
\node[anchor=west] (B) at (0,-1.3) {$(U(N)_k\times U(N)_{-k})/\mathbb{Z}_{m'}$};
\node[anchor=west] (C) at (0,-2.6) {$\phantom{(}U(N)_k\times U(N)_{-k}$};
\draw[->] (A.south) -- node [right] {$\mathbb{Z}_{m}^{[0]}$}  (A.south|-B.north);
\draw[->] (A.south|-B.south) --  node [right] {$\mathbb{Z}_{m'}^{[0]}$} (A.south|-C.north);
\draw[->] (C.north) --node [left] {$\mathbb{Z}_{m'}^{[1]}$} (C.north|-B.south);
\draw[->] (C.north|-B.north) -- node [left] {$\mathbb{Z}_{m}^{[1]}$}(C.north|-A.south);

\node (X) at (3.7,0) {$=$};

\node[anchor=west] (P) at (4,0) {$(SU(N)_k\times SU(N)_{-k})/\mathbb{Z}_N$};
\node[anchor=west] (Q) at (4,-1.3) {$(SU(N)_k\times SU(N)_{-k})/\mathbb{Z}_{n'}$};
\node[anchor=west] (R) at (4,-2.6) {$\phantom{(}SU(N)_k\times SU(N)_{-k}$};
\draw[->] (P.south) -- node [right] {$\mathbb{Z}_{n}^{[0]}$}  (P.south|-Q.north);
\draw[->] (P.south|-Q.south) --  node [right] {$\mathbb{Z}_{n'}^{[0]}$} (P.south|-R.north);
\draw[->] (R.north) --node [left] {$\mathbb{Z}_{n'}^{[1]}$} (R.north|-Q.south);
\draw[->] (R.north|-Q.north) -- node [left] {$\mathbb{Z}_{n}^{[1]}$}(R.north|-P.south);
\end{tikzpicture}
\caption{The general discrete gauging relations for ${\cal N}=6$ SCS theories.}
\label{DiscreteGauging2}
\end{figure}

The properties of both sets of theories, $(U(N)_k\times U(N)_{-k})/\mathbb{Z}_{m'}$ and $(SU(N)_k\times SU(N)_{-k})/\mathbb{Z}_{n'}$,
are summarized in Table~\ref{FieldTheoryProperties}. We denote by ${\cal B}^\ell$ an $\ell$-fold product of the minimal 
dressed di-baryon operator,
by ${\cal M}_\ell$ a dressed monopole operator corresponding to $\ell$ units of the magnetic flux sourced by the minimal 
dressed monopole,
and by ${\cal W}_\ell$ a Wilson line in the $\ell$-fold product of the $({\bf N},{\bf N})$ representation.

\begin{table}[h!]
\begin{center}
\begin{tabular}{|l|l|l|l|}
 \hline 
 theory & global symmetry & spectrum & charges \\
\hline
$(U(N)_k\times U(N)_{-k})/\mathbb{Z}_{m'}$  & $U[1]^{[0]}\times \mathbb{Z}_{\gcd(N,m')}^{[0]}$   & ${\cal M}_\ell$ & $(\ell,0)$ \\
\qquad where $k=mm'$ && ${\cal B}^{m\ell}$ & $\left(-\frac{m\ell N}{k},\ell \, \mbox{mod}\, \gcd(N,m')\right)$\\ 
& $\mathbb{Z}_{m}^{[1]}$ & ${\cal W}_{m'\ell}$ & $\ell \, \mbox{mod} \, m$ \\ 
 \hline 
$(SU(N)_k\times SU(N)_{-k})/\mathbb{Z}_{n'}$   & $U[1]^{[0]}\times \mathbb{Z}_{\gcd(n',k)}^{[0]}$ & ${\cal B}_\ell$ & $(\ell,0)$ \\
\qquad where $N=nn'$ && ${\cal M}_{n\ell}$ & $\left(-\frac{n\ell k}{N},\ell \, \mbox{mod} \, \gcd(n',k)\right)$ \\ 
& $\mathbb{Z}_{n}^{[1]}$ & ${\cal W}_{n'\ell}$ & $\ell\, \mbox{mod} \, n$ \\
 \hline
 \end{tabular}
 \end{center}
\caption{Global symmetries and charge spectrum of ${\cal N}=6$ SCS theories.}
\label{FieldTheoryProperties}
\end{table}

\section{AdS/CFT with boundary conditions}

Next we will determine how all the theories in Table~\ref{FieldTheoryProperties} fit into the dual supergravity description.

\subsection{Review of the basics}

As argued in \cite{Aharony:2008ug}, the $U(N)_k\times U(N)_{-k}$ theory is dual to M-theory on $AdS_4\times S^7/\mathbb{Z}_k$,
or equivalently to Type IIA string theory on $AdS_4\times \mathbb{C}P^3$.
Let us briefly recall the relevant details of the Type IIA description, which is the one that 
will be more convenient for our purpose.
The Type IIA string theory background has RR fluxes given by
\be
\label{RRfluxes}
\int_{\mathbb{C}P^3}\frac{F_6}{2\pi} \sim N \,, \qquad \int_{\mathbb{C}P^1\subset \mathbb{C}P^3}\frac{F_2}{2\pi} \sim k \,,
\ee
corresponding, respectively, to the rank and the CS level of the gauge theory.
Upon reduction on $\mathbb{C}P^3$ we can identify three Abelian gauge fields in $AdS_4$.
The first is the NSNS two-form $B$ that couples electrically to fundamental strings.
This field will be related to the one-form symmetry of the gauge theory.
The other two gauge fields are both one-forms, and are given by the RR one-form $C_1$ and by 
the reduction of the RR three-form $C_3$ on the $\mathbb{C}P^1$ two-cycle inside $\mathbb{C}P^3$.
These couple electrically to D0-branes and to D2-branes wrapping $\mathbb{C}P^1$, respectively.
It is actually more convenient in the latter case to work with the 4d magnetic dual gauge field
that couples electrically to D4-branes wrapping $\mathbb{C}P^2\subset \mathbb{C}P^3$.
We will therefore denote the two one-form gauge fields as $A_{D0}$ and $A_{D4}$, respectively.
There are also magnetically charged objects: 
a D6-brane wrapped on $\mathbb{C}P^3$ is charged magnetically under $A_{D0}$, and a D2-brane wrapped on $\mathbb{C}P^1$
is charged magnetically under $A_{D4}$.
However these two objects come with strings attached due to worldvolume tadpoles induced by the RR fluxes (\ref{RRfluxes}):
the wrapped D6-brane has $N$ strings attached, and the wrapped D2-brane has $k$ strings attached.

The one-form gauge fields in $AdS_4$ should be related to the zero-form symmetry of the gauge theory.
However, as was already noted in \cite{Aharony:2008ug}, only one combination of the two one-form gauge fields is  massless.
Here we note that this is directly related to the following topological term in the 4d effective action,
\be
\label{IIAAction}
S_{top} &=& \frac{1}{2\pi} \int_{AdS_4} B\wedge d(NA_{D0} + kA_{D4}).
\ee
This represents a St\"uckelberg-like term for the combination $NA_{D0} + kA_{D4}$,
where the role of the would-be Goldstone boson is played by the magnetic dual of the two-form $B$.
This means that the zero-form gauge symmetry in the bulk is spontaneously broken to $U(1)\times \bZ_{\gcd(N,k)}\subset U(1)_{D0}\times U(1)_{D4}$.
The massless $U(1)$ gauge field $A$ is parametrized as 
\be
\label{MasslessA}
(A_{D0},A_{D4})=(-qA, pA) \,,
\ee 
where we recall that the co-prime integers $p, q$ were defined by $N=p\gcd(N,k)$ and $k=q\gcd(N,k)$.
The D0-brane and the wrapped D4-brane are both charged under the unbroken $U(1)$ gauge symmetry,
with the ratio of their charges given by $q/p = k/N$.
This leads us to identify $A_{D0}$ and $A_{D4}$ as the bulk gauge fields dual to the $U(1)_{\cal M}$ and $U(1)_{\cal B}$ symmetries, respectively,
and correspondingly to identify the D0-brane and the wrapped D4-brane as the bulk states dual to the dressed-monopole operator ${\cal M}_1$
and the di-baryon operator ${\cal B}$, respectively.
The masses and dimensions also agree, since
\be
\Delta_{\cal M} = m_{D0} R \sim \frac{R}{g_s \ell_s} \sim k/2 \,,
\ee
and 
\be
\Delta_{\cal B} = m_{D4} R \sim T_{D4} R^5 \sim \frac{R^5}{g_s \ell_s^5} \sim \frac{kR^4}{\ell_s^4} \sim N/2 \,.
\ee

But this is puzzling given that this $AdS_4$ background was originally found as the dual of the $U(N)_k\times U(N)_{-k}$ theory,
which as we have explained does not have a gauge invariant di-baryon operator.
As we will soon see, the resolution of this puzzle lies in understanding the boundary conditions.

But before we discuss the boundary conditions,
let us first be more precise about the meaning of the topological term 
\eqref{IIAAction} \emph{in the bulk}.
This is the dominant term for the gauge fields in the 4d low energy effective theory near the boundary of $AdS_4$.
The equations of motion that follow from this action are 
\be
NdB &=& 0 \\
kdB &=& 0 \\
d(NA_{D0}+kA_{D4})&=& \gcd(N,k)d(pA_{D0} + qA_{D4}) = 0\,.
\ee
The first two equations imply that $B$ is a $\bZ_{\gcd(N,k)}$-valued two-form gauge field,
and the third one implies that the combination $pA_{D0} + q A_{D4}$ is a $\bZ_{\gcd(N,k)}$-valued one-form gauge field.
The orthogonal combination given by $A$ as defined in (\ref{MasslessA}) remains as a $U(1)$ one-form gauge field.
This is basically what we observed above.

\subsection{``Standard" boundary conditions}

We begin with the  ``standard" set of boundary conditions fixing the values of the one-form gauge fields $A_{D0}$ and $A_{D4}$ on the boundary.
In other words $A_{D0}$ and $A_{D4}$ satisfy Dirichlet boundary conditions.
We can then allow the two-form gauge field $B$ to be free on the boundary, but 
the boundary values of $A_{D0}$ and $A_{D4}$ need to be compatible with this. 
Due to the topological term (\ref{IIAAction}), the boundary values of $A_{D0}$ and $A_{D4}$ must satisfy $NA_{D0} + kA_{D4}=0$.
This means that the background gauge field one can specify at the boundary is $G_0\sim U(1)\times \mathbb{Z}_{\gcd(N,k)}$.
In other words the boundary theory is the $(SU(N)_k\times SU(N)_{-k})/\mathbb{Z}_N$, 
or equivalently $(U(N)_k\times U(N)_{-k})/\mathbb{Z}_k$ theory.

\begin{figure}[h!]
\center
\includegraphics[height=0.2\textwidth]{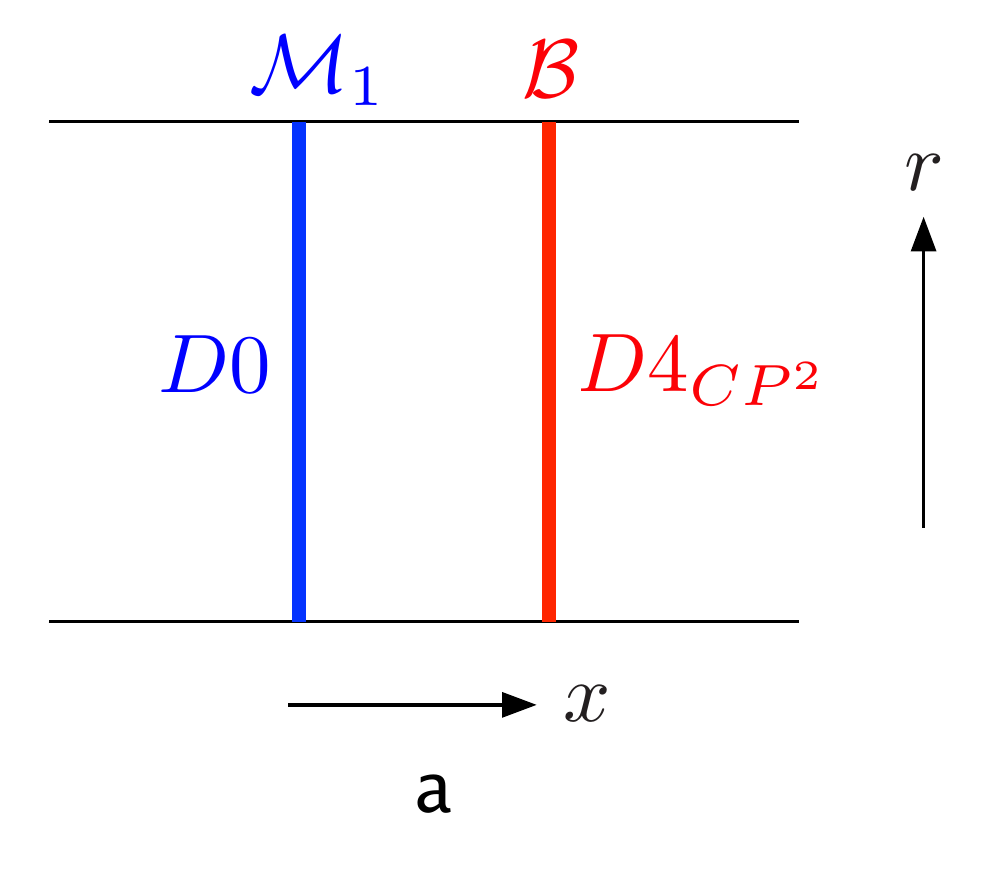} 
\hspace{20pt}
\includegraphics[height=0.2\textwidth]{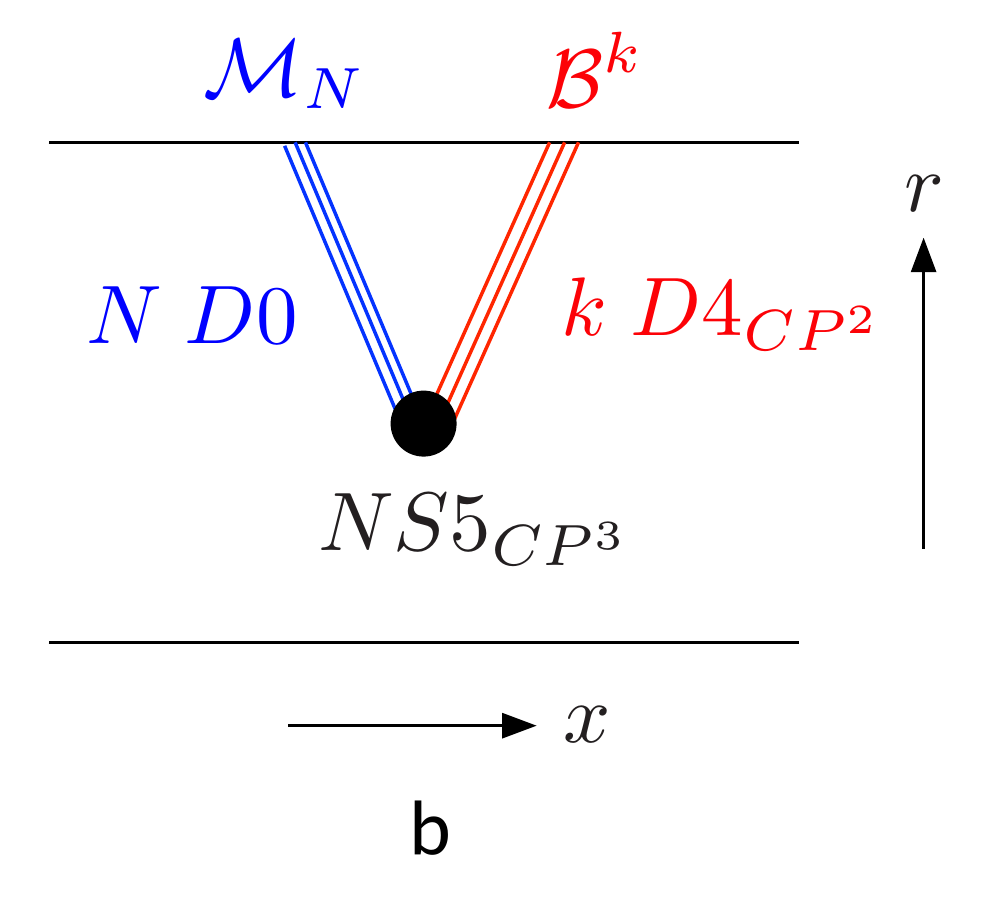} 
\caption{A holographic description of (a) a dressed monopole and a  di-baryon, and (b) the relation ${\cal B}^k = {\cal M}_{-N}$,
in the $(U(N)_k\times U(N)_{-k})/\mathbb{Z}_k = (SU(N)_k\times SU(N)_{-k})/\mathbb{Z}_N$ theory.}
\label{AdS4-1}
\end{figure}

We can understand this more concretely as follows.
On the one hand, the free boundary condition for $B$ forbids strings from ending on the boundary, 
and therefore the boundary theory has no unscreened Wilson lines.
On the other hand the boundary conditions for $A_{D0}$ and $A_{D4}$ mean that 
D0-branes and wrapped D4-branes are allowed to end on the boundary.
The boundary theory should therefore have two types of local operators charged under the global $U(1)$ symmetry, 
with a charge ratio $k/N$.
These are the dressed monopole and di-baryon operators, see Fig.~\ref{AdS4-1}a.
Furthermore, $N$ D0-branes can turn into $k$ wrapped D4-branes via an instantonic NS5-brane wrapped on $\mathbb{C}P^3$,
realizing the chiral-ring-like relation (\ref{ChiralRing}) between the di-baryon and dressed monopole, see Fig.~\ref{AdS4-1}b.\footnote{One way to see this is by going to the 
M-theory description, in which the $N$ D0-branes become $Nk$ units of momentum on $S^7$,
which can become a maximal giant M5-brane, which in turn maps to $k$ D4-branes wrapping $\mathbb{C}P^2$ \cite{Aharony:2008ug}.
Another way to see this is from the worldvolume theory of the fully wrapped NS5-brane, in which the worldvolume scalar potential has an electric tadpole of size $N$ due
to the RR 6-form flux on $\mathbb{C}P^3$, and a magnetic tadpole of size $k$ due to the RR two-form flux on $\mathbb{C}P^1$.
The former is cancelled by having $N$ D0-brane worldlines end on the NS5-brane, and the latter is cancelled by having $k$ wrapped D4-brane
worldlines end on it.}
All of this is consistent with the identification of the boundary theory as the 
$(SU(N)_k\times SU(N)_{-k})/\mathbb{Z}_N = (U(N)_k\times U(N)_{-k})/\mathbb{Z}_k$ theory.

\subsection{``Alternative" boundary conditions}

The fact that the ``standard" boundary conditions that fix the boundary values of both one-form fields correspond to 
the $(SU(N)_k\times SU(N)_{-k})/\mathbb{Z}_N = (U(N)_k\times U(N)_{-k})/\mathbb{Z}_k$ theory mirrors
the fact that this is in some sense the most ``basic" ${\cal N}=6$ SCS theory, from which all other theories can be obtained by gauging
a discrete subgroup of the global zero-form symmetry.
From the bulk viewpoint, all other ${\cal N}=6$ theories will correspond to changing the boundary conditions for $B, A_{D0}$ and $A_{D4}$
in a way that is consistent with the topological term \eqref{IIAAction}.
We will not attempt to classify all allowed boundary conditions.
But we will find the boundary conditions that correspond to the set of ${\cal N}=6$ SCS theories discussed in \cite{Tachikawa:2019dvq}.
In particular we will identify the boundary conditions dual to the $U(N)_k\times U(N)_{-k}$ theory, which will allow
us to resolve the di-baryon puzzle.

\subsubsection{$U(N)_k\times U(N)_{-k}$}

To get the $U(N)_k\times U(N)_{-k}$ theory, we fix the boundary value of $A_{D0}$, 
but allow the boundary value of $A_{D4}$ to be free.\footnote{It may be possible to allow both
 $A_{D0}$ and $A_{D4}$ to be free at the boundary, but we will not consider that possibility here.
 This corresponds to performing Witten's $SL(2,\mathbb{Z})$ operation \cite{Witten:2003ya} on the ABJM theory, and presumably is not compatible with $\mathcal{N}=6$ supersymmetry.
 }
The boundary theory therefore has a $U(1)_{\cal M}$ global zero-form symmetry, but the $U(1)_{\cal B}$ symmetry is gauged.
More precisely, the boundary values of $A_{D4}$ are free to fluctuate in $\mathbb{Z}_k$,
which in essence means that the $\mathbb{Z}_k$ subgroup of $G_0\subset U(1)_{\cal M}\times U(1)_{\cal B}$, the global symmetry in the case of the ``standard" boundary conditions, is gauged.
Due to the free boundary condition on $A_{D4}$, we cannot take the boundary value of $B$ to be free.
The coupling $\frac{k}{2\pi}B\wedge dA_{D4}$ requires the boundary value of $B$ to be fixed, such that its holonomy takes a boundary value
in $\mathbb{Z}_k$, namely $k \int_{S^2} B|_{\partial} =0$ mod $2\pi$.
With a slight abuse of notation we will denote the boundary holonomy of $B$ simply by $B$, so the boundary condition is $kB=0$ mod $2\pi$.
The boundary theory therefore also has a global $\mathbb{Z}_k$ one-form symmetry.
These are precisely the global symmetries of the $U(N)_k\times U(N)_{-k}$ theory.
The remaining bulk coupling $\frac{N}{2\pi}  B \wedge dA_{D0}$
is identified with the 4d characteristic class corresponding to the 3d mixed anomaly \eqref{Zk-anomaly},
upon identifying $B=(2\pi/k)\mathsf{B}$ and $A_{D0}=A$.

\begin{figure}[h!]
\center
\includegraphics[height=0.2\textwidth]{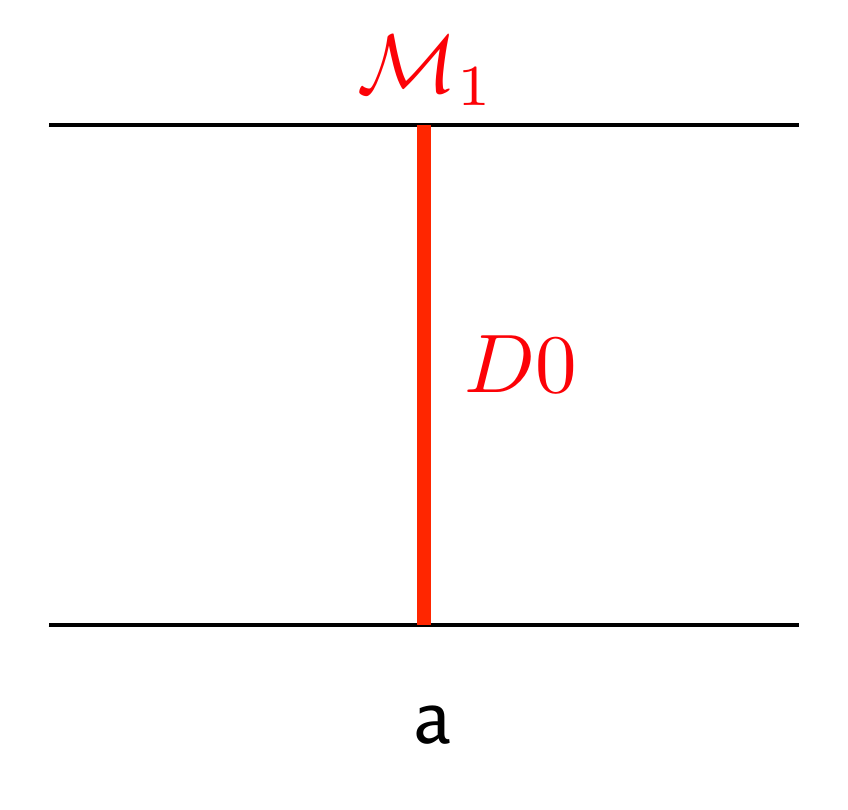} 
\hspace{20pt}
\includegraphics[height=0.2\textwidth]{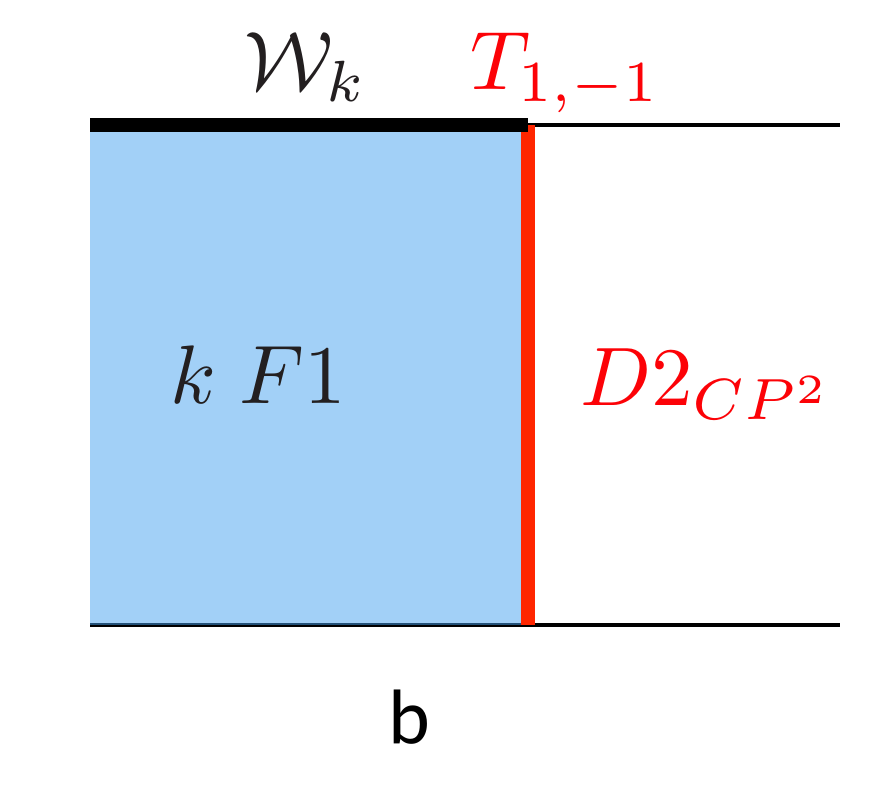} 
\caption{A holographic description of (a) a dressed monopole, and (b) $k$ Wilson lines being screened by an
antisymmetric monopole in the $U(N)_k\times U(N)_{-k}$ theory.}
\label{AdS4-2}
\end{figure}

In terms of branes, D0-branes are allowed to end on the boundary of $AdS_4$, but wrapped D4-branes are not.
This agrees with what we know about the $U(N)_k\times U(N)_{-k}$ theory.
The boundary gauge theory has dressed monopole operators corresponding to the endpoints of D0-brane worldlines,
but does not have a di-baryon operator which would correspond to the endpoint of a wrapped D4-brane worldline.
In addition, the boundary condition for $B$ allows a fundamental string worldsheet to end on the boundary, and the resulting boundary line
corresponds to the $({\bf N},{\bf N})$ Wilson line of the gauge theory.
A collection of $k$ such strings can end on a wrapped D2-brane, which, being the magnetic dual of the wrapped D4-brane, 
is allowed to end on the boundary.
This is the bulk description of the $T_{1,-1}$ monopole screening a $k$-fold product of the basic Wilson line.
Finally, the wrapped D6-brane is not allowed to end on the boundary since the D0-brane is, so there is not an additional $N$-fold screening 
of the Wilson lines.
See Figure~\ref{AdS4-2} for illustrations.

\subsubsection{$SU(N)_k\times SU(N)_{-k}$}
The $SU(N)_k\times SU(N)_{-k}$ theory corresponds to exchanging the roles of $A_{D0}$ and $A_{D4}$.
Namely, we fix the boundary value of $A_{D4}$, but allow the boundary value of $A_{D0}$ to be free.
In this case the boundary theory has a $U(1)_{\cal B}$ global zero-form symmetry, but the $U(1)_{\cal M}$ symmetry is gauged.
More precisely, the boundary values of $A_{D0}$ are free to fluctuate in $\mathbb{Z}_N$, which in essence means that the 
the $\mathbb{Z}_N$ subgroup of $G_0\subset U(1)_{\cal M}\times U(1)_{\cal B}$ is gauged.
Due to the coupling $\frac{N}{2\pi}B\wedge dA_{D0}$, the free boundary condition for $A_{D0}$ requires the boundary value of $B$
to be fixed to a value in $\mathbb{Z}_N$, {\em i.e.} $NB=0$ mod $2\pi$ (using the same abuse of notation as before).
The boundary theory therefore also has a global $\mathbb{Z}_N$ one-form symmetry.
These are precisely the global symmetries of the $SU(N)_k\times SU(N)_{-k}$ theory.
The remaining bulk coupling $\frac{k}{2\pi}  B \wedge dA_{D4}$
is identified with the 4d characteristic class corresponding to the 3d mixed anomaly \eqref{ZN-anomaly},
upon identifying $B=(2\pi/N)\mathsf{B}$ and $A_{D4}=A$.

\begin{figure}[h!]
\center
\includegraphics[height=0.2\textwidth]{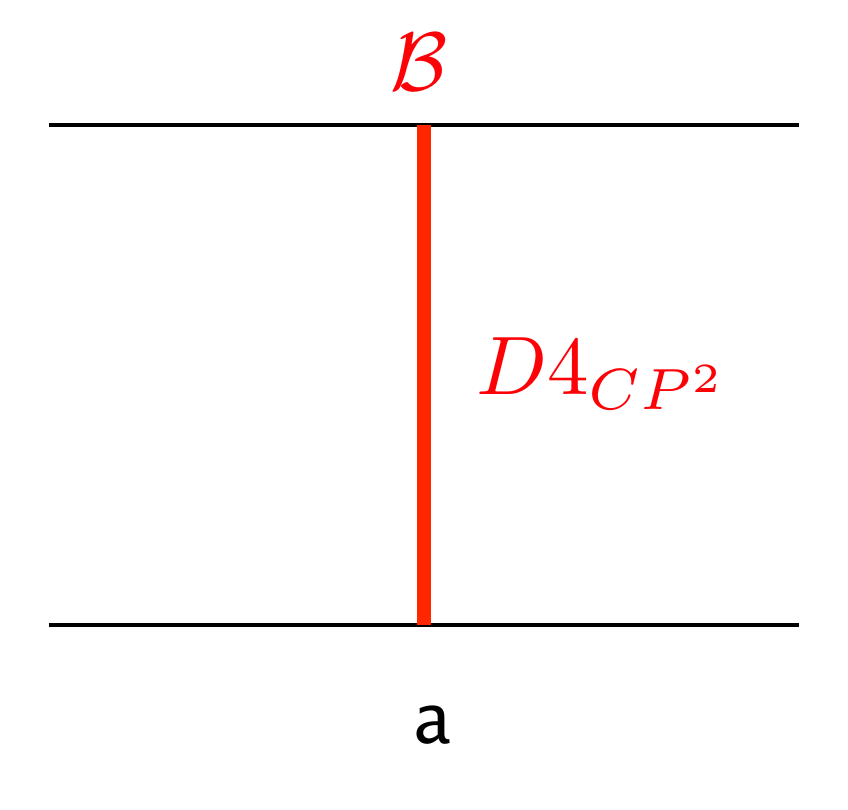} 
\hspace{20pt}
\includegraphics[height=0.2\textwidth]{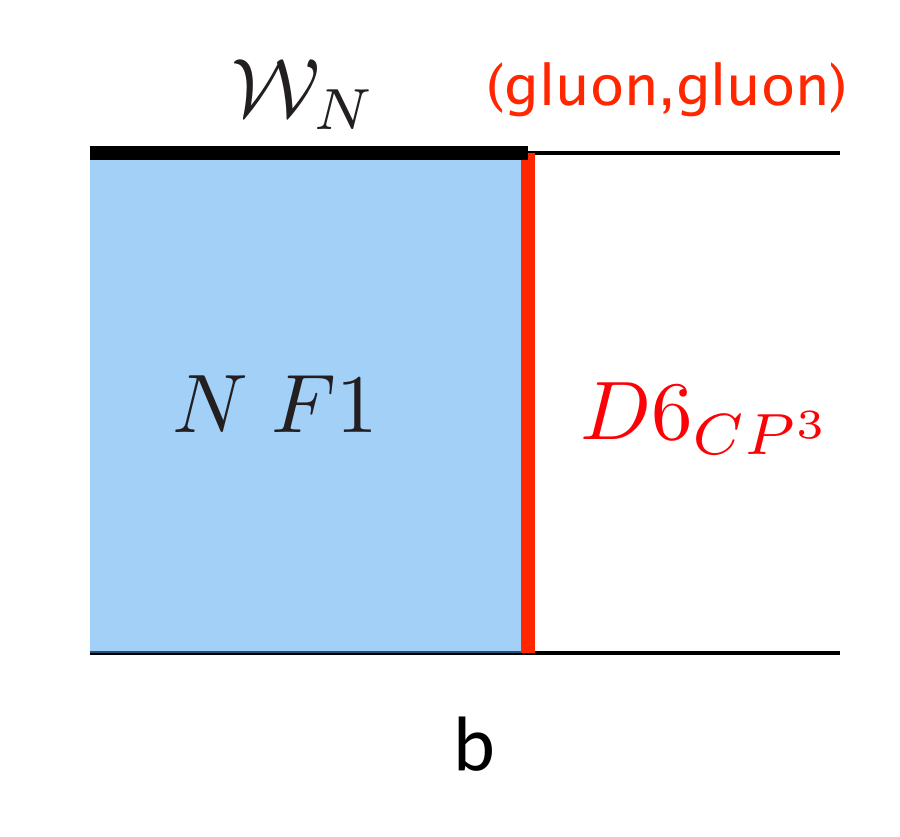} 
\caption{A holographic description of (a) a baryon, and (b) $N$ Wilson lines being screened by 
gluons in the $SU(N)_k\times SU(N)_{-k}$ theory.}
\label{AdS4-3}
\end{figure}

Now wrapped D4-branes are allowed to end on the boundary whereas D0-branes are not.
This agrees with what we know about the $SU(N)_k\times SU(N)_{-k}$ theory.
The boundary gauge theory has a di-baryon operator corresponding to the endpoint of a wrapped D4-brane worldline,
but does not have monopole operators which would correspond to the endpoints of D0-brane worldlines.
The boundary condition for $B$ again allows a fundamental string worldsheet to end on the boundary, and the resulting boundary line
corresponds to the $({\bf N},{\bf N})$ Wilson line of the gauge theory.
Now a collection of $N$ such strings can end on a wrapped D6-brane, which, being the magnetic dual of the D0-brane, 
is allowed to end on the boundary.
This is the bulk description of the gluon screening an $N$-fold product of the basic Wilson line.
 Finally, the wrapped D2-brane is not allowed to end on the boundary since the wrapped D4-brane is, so there is not an additional $k$-fold screening 
of the Wilson lines.
See Figure~\ref{AdS4-3} for illustrations.

\subsection{Generalization}

In the two ``alternative" boundary conditions we discussed above, we fixed one of the one-form gauge fields at the boundary,
while keeping the other one maximally free, within the discrete symmetry imposed by the action (\ref{IIAAction}), $\mathbb{Z}_k$ or $\mathbb{Z}_N$.
This in turn required fixing the boundary condition for the two-form gauge field $B$ to take a value in this group.
If either $\mathbb{Z}_k$ or $\mathbb{Z}_N$ have a non-trivial subgroup there is a natural way to generalize these boundary conditions,
by partially restricting the freedom of the free one-form gauge field to this subgroup, 
which allows us at the same time to partially loosen the restriction on the two-form gauge field, 
giving it freedom within the complement of this subgroup.
In either case we will keep the Dirichlet boundary condition for the other one-form gauge field.\footnote{In principle there is a more
general possibility of restricted free boundary conditions on both one-form gauge fields. We will not consider that here.}
Our results are summarized in Table~\ref{Dualities} below, and the details are contained in the following two subsections.

\begin{table}[h!]
\begin{center}
\begin{tabular}{|l|l|}
 \hline 
$AdS_4$ boundary conditions & {3d ${\cal N}=6$ theory} \\
 \hline 
 $A_{D0}$ fixed & $(U(N)_k\times U(N)_{-k})/\mathbb{Z}_{m'}$ \\
 $A_{D4}$ free mod $\mathbb{Z}_{m'}\subset \mathbb{Z}_k$ & where $k=mm'$ \\
 $B$ free mod $\mathbb{Z}_{m}\subset \mathbb{Z}_k$ & \\
\hline
$A_{D4}$ fixed &  $(SU(N)_{k}\times SU(N)_{-k})/\mathbb{Z}_{n'}$ \\
$A_{D0}$ free mod $\mathbb{Z}_{n'}\subset \mathbb{Z}_N$ & where $N=nn'$\\
$B$ free mod $\mathbb{Z}_{n}\subset \mathbb{Z}_N$ & \\
\hline
 \end{tabular}
 \end{center}
\caption{Generalized $AdS_4/CFT_3$ dualities.}
\label{Dualities}
\end{table}

\subsubsection{$(U(N)_k\times U(N)_{-k})/\mathbb{Z}_{m'}$}

If $k=mm'$ we can restrict the boundary value of $A_{D4}$ to be free
within $\mathbb{Z}_m \subset \mathbb{Z}_k$, while fixing the boundary value of $A_{D0}$.
We can also say that $A_{D4}$ is free in $\mathbb{Z}_k$ modulo fixing it in $\mathbb{Z}_{m'}\subset \mathbb{Z}_k$, namely 
$m'A_{D4}=0$.\footnote{More explicitly, $A_{D4}$ is allowed to vary in the set $\{0,m',2m',\ldots,k-m'\}$ or
$\{1,m'+1,2m'+1,\ldots,k-m'+1\}$ or $\{2,m'+2,2m'+2,\ldots,k-m'+2\}$,..., or $\{m'-1,2m'-1,3m'-1,\ldots,k-1\}$.}
From the point of view of the boundary theory we are gauging $\mathbb{Z}_m\subset \mathbb{Z}_k\subset G_0$,
leaving a discrete zero-form global symmetry $\mathbb{Z}_{\gcd(N,m')}$, in addition to the $U(1)_{\cal M}$ global zero-form symmetry dual to $A_{D0}$.
The restriction on the boundary freedom of $A_{D4}$ in turn allows us to relax the boundary condition for $B$,
giving it freedom in $\mathbb{Z}_{m'} \subset \mathbb{Z}_k$.
We can say that $B$ is free in $\mathbb{Z}_k$ modulo fixing it in $\mathbb{Z}_{m}\subset \mathbb{Z}_k$, namely $mB=0$.
This gives rise to a $\mathbb{Z}_m$ global one-form symmetry.
The full global symmetry of the boundary theory is therefore $U(1)_{\cal M}^{[0]} \times \mathbb{Z}_{\gcd(N,m')}^{[0]} \times \mathbb{Z}_m^{[1]}$,
which we recognize as the symmetry of the $(U(N)_k\times U(N)_{-k})/\mathbb{Z}_{m'}$ theory.
The above boundary condition interpolates between the ``standard" boundary condition for $(m,m')=(1,k)$ and the $U(N)_k\times U(N)_{-k}$ 
boundary condition for $(m,m')=(k,1)$.
As in the previous cases, the bulk coupling should reproduce the mixed anomaly expected between the one-form and zero-form symmetries.

\begin{figure}[h!]
\center
\includegraphics[height=0.2\textwidth]{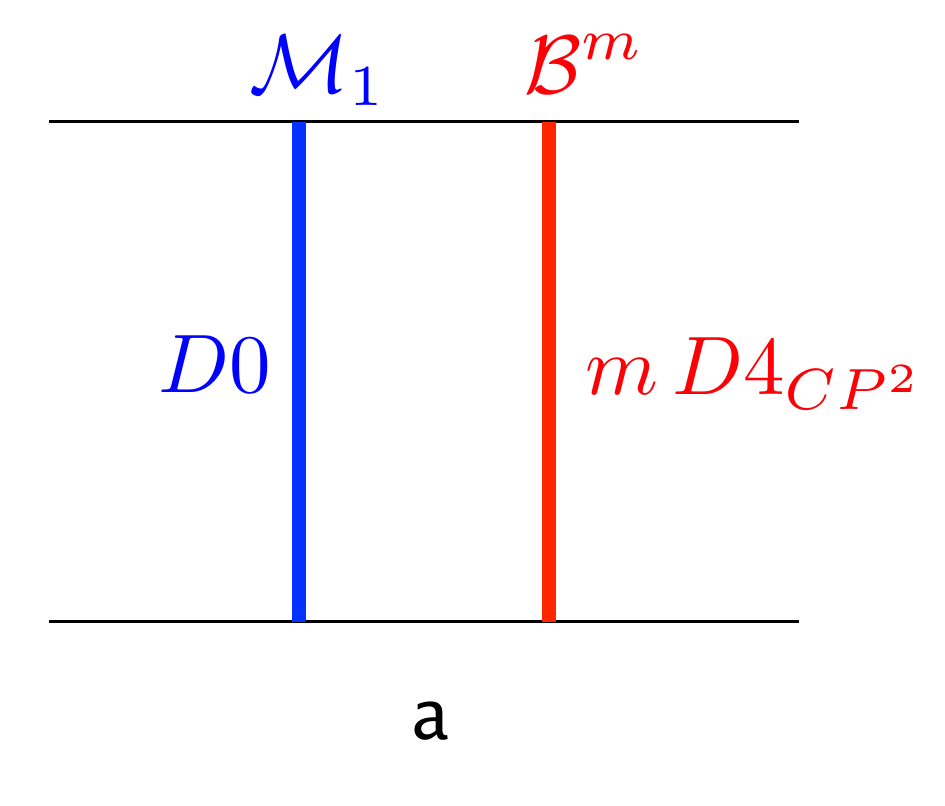} 
\hspace{20pt}
\includegraphics[height=0.2\textwidth]{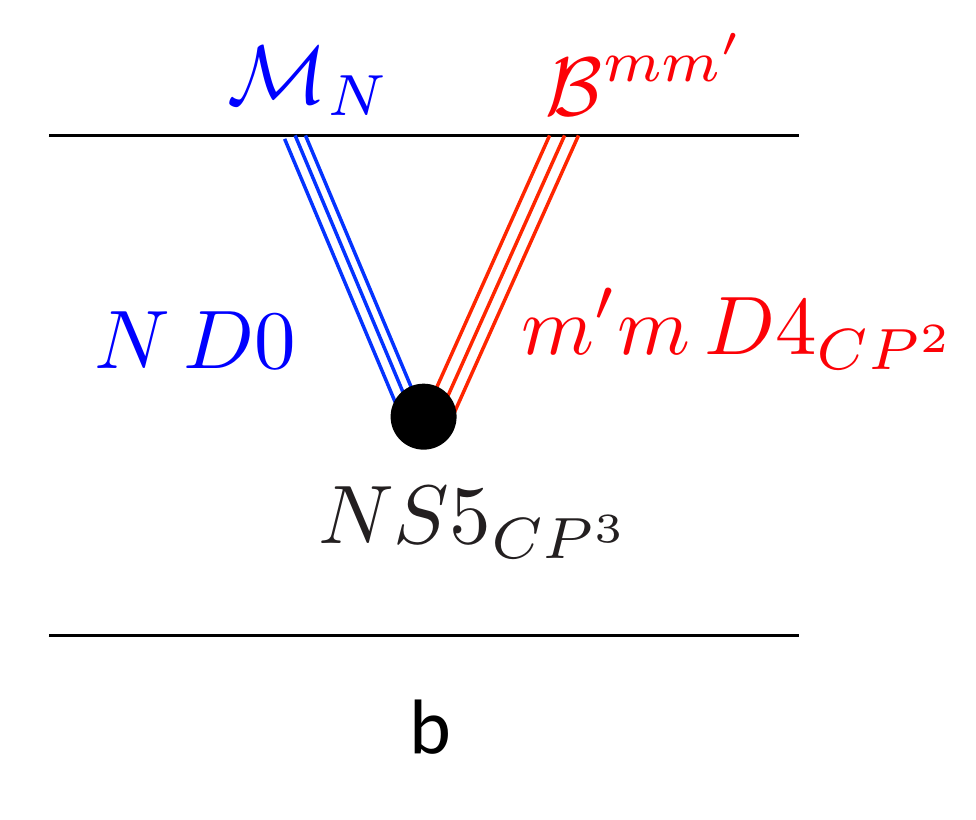} 
\hspace{20pt}
\includegraphics[height=0.2\textwidth]{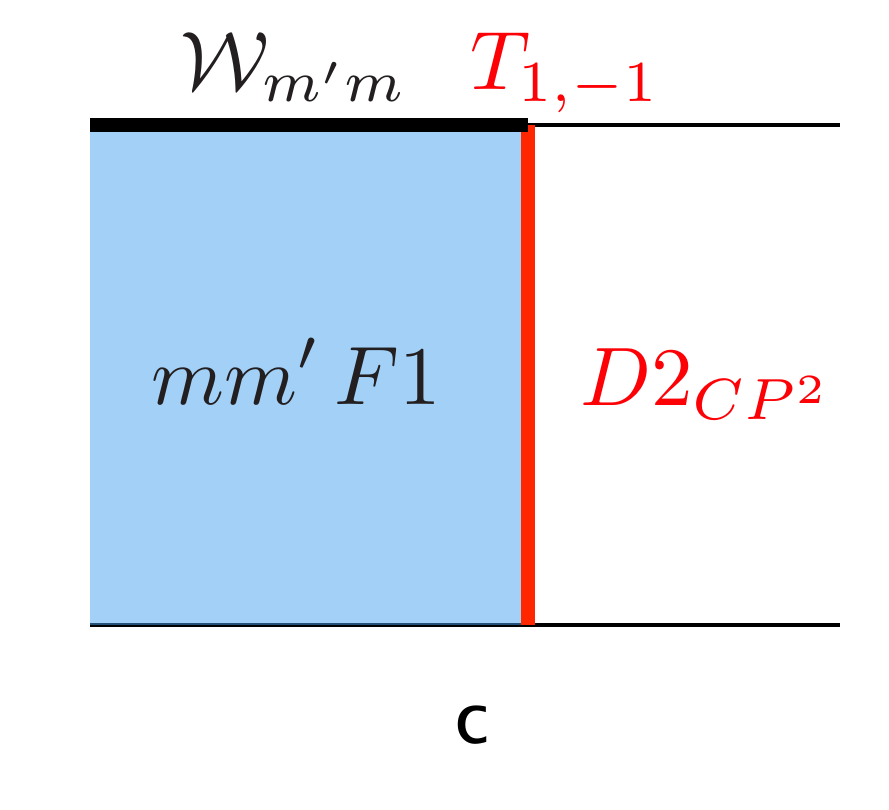} 
\caption{$AdS_4$ dual of $(U(N)_k\times U(N)_{-k})/\mathbb{Z}_{m'}$: (a) Dressed-monopole and $m$-di-baryon operators. (b) Chiral ring relation. (c) $k=m'm$ fundamental Wilson lines screened by an anti-symmetric monopole.}
\label{AdS4-4}
\end{figure}

The Dirichlet boundary condition for $A_{D0}$ allows D0-branes to end on the boundary, giving the dressed monopole operators ${\cal M}_\ell$.
The $\mathbb{Z}_m$ restricted-free, or equivalently the $\mathbb{Z}_{m'}$ fixed, boundary condition for $A_{D4}$,
allows also wrapped D4-branes to end on the boundary in multiples of $m$. These correspond to $m$-fold products of the di-baryon operator
${\cal B}^{m\ell}$.
The chiral ring relation, as in the case with the ``standard" boundary conditions, is described in the bulk as a fully wrapped Euclidean NS5-brane.
The $\mathbb{Z}_m$ fixed boundary condition for $B$ allows string worldsheets to end on the boundary in multiples of $m'$,
describing the Wilson lines ${\cal W}_{m'\ell}$,
with $m$ of these multiples screened by a wrapped D2-brane.
See Fig.~\ref{AdS4-4} for illustrations.
All of this agrees with the properties of $(U(N)_k\times U(N)_{-k})/\mathbb{Z}_{m'}$ theory shown in Table~\ref{FieldTheoryProperties}.

\subsubsection{$(SU(N)_k\times SU(N)_{-k})/\mathbb{Z}_{n'}$}

If $N=nn'$ we can restrict the boundary value of $A_{D0}$ to be free
within $\mathbb{Z}_n \subset \mathbb{Z}_N$, while fixing the boundary value of $A_{D4}$.
In other words $A_{D0}$ is free in $\mathbb{Z}_N$ modulo fixing it in $\mathbb{Z}_{n'}\subset \mathbb{Z}_N$, namely 
$n'A_{D0}=0$.
From the point of view of the boundary theory we are gauging $\mathbb{Z}_n\subset \mathbb{Z}_N\subset G_0$,
leaving a discrete zero-form global symmetry $\mathbb{Z}_{\gcd(n',k)}$, in addition to the $U(1)_{\cal B}$ global zero-form symmetry dual to $A_{D4}$.
The restriction on the boundary freedom of $A_{D0}$ in turn allows us to relax the boundary condition for $B$,
giving it freedom in $\mathbb{Z}_{n'} \subset \mathbb{Z}_N$.
We can say that $B$ is free in $\mathbb{Z}_N$ modulo fixing it in $\mathbb{Z}_{n}\subset \mathbb{Z}_N$, namely $nB=0$.
This gives rise to a $\mathbb{Z}_n$ global one-form symmetry.
The full global symmetry of the boundary theory is therefore $U(1)_{\cal B}^{[0]} \times \mathbb{Z}_{\gcd(n',k)}^{[0]} \times \mathbb{Z}_n^{[1]}$,
which we recognize as the symmetry of the $(SU(N)_k\times SU(N)_{-k})/\mathbb{Z}_{n'}$ theory.
The above boundary condition interpolates between the ``standard" boundary condition for $(n,n')=(1,N)$ and the $SU(N)_k\times SU(N)_{-k}$ 
boundary condition for $(n,n')=(N,1)$.
The bulk coupling should again reproduce the mixed anomaly expected between the one-form and zero-form symmetries.

\begin{figure}[h!]
\center
\includegraphics[height=0.2\textwidth]{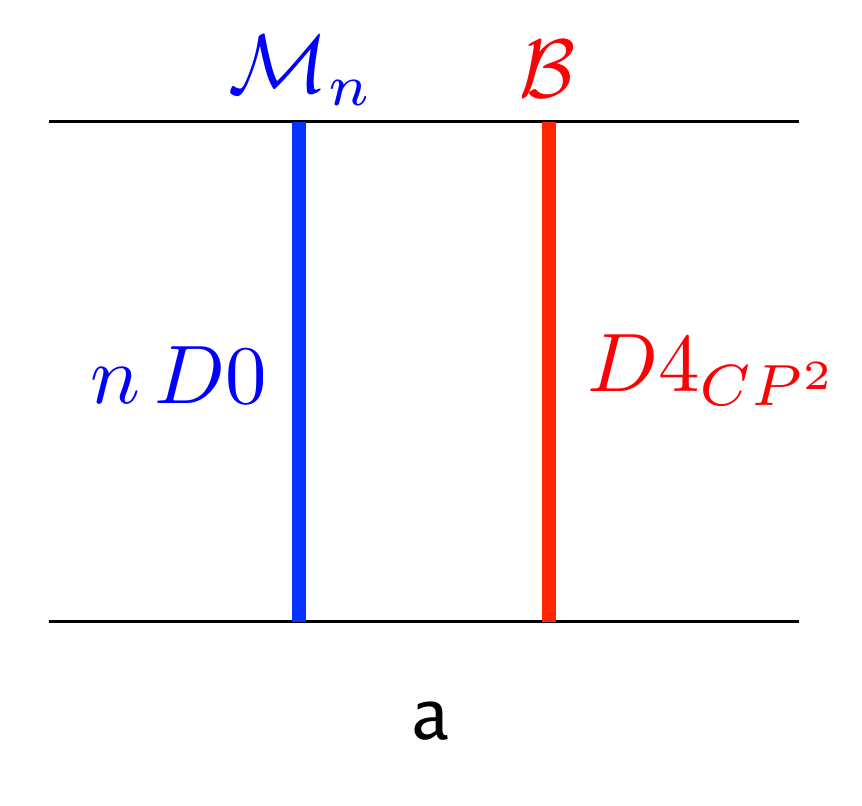} 
\hspace{20pt}
\includegraphics[height=0.2\textwidth]{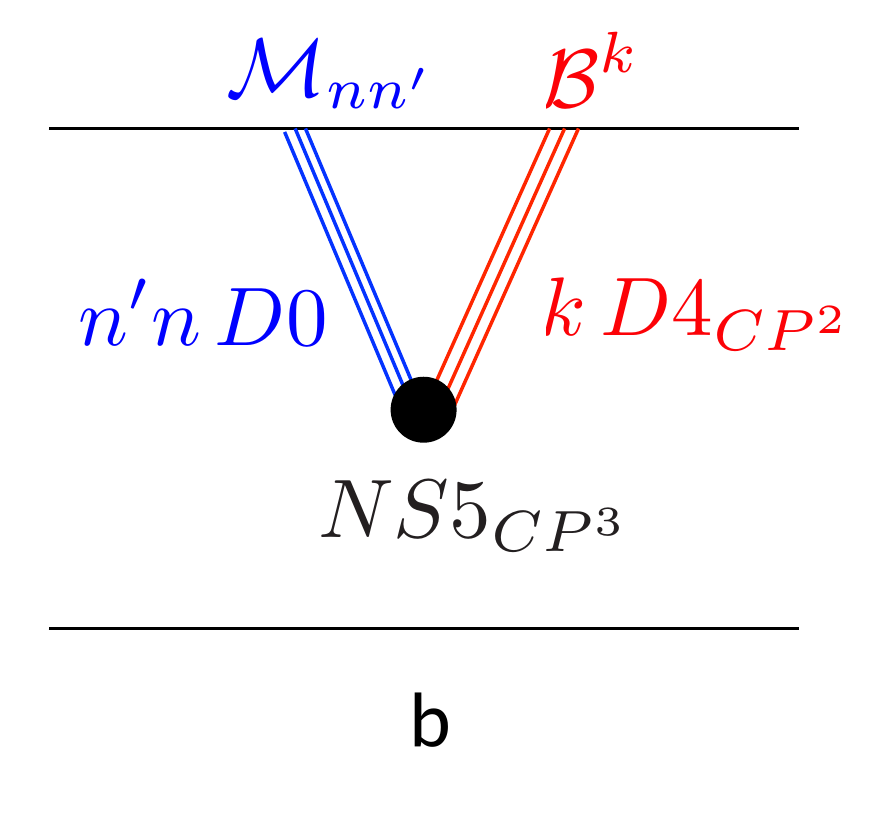} 
\hspace{20pt}
\includegraphics[height=0.2\textwidth]{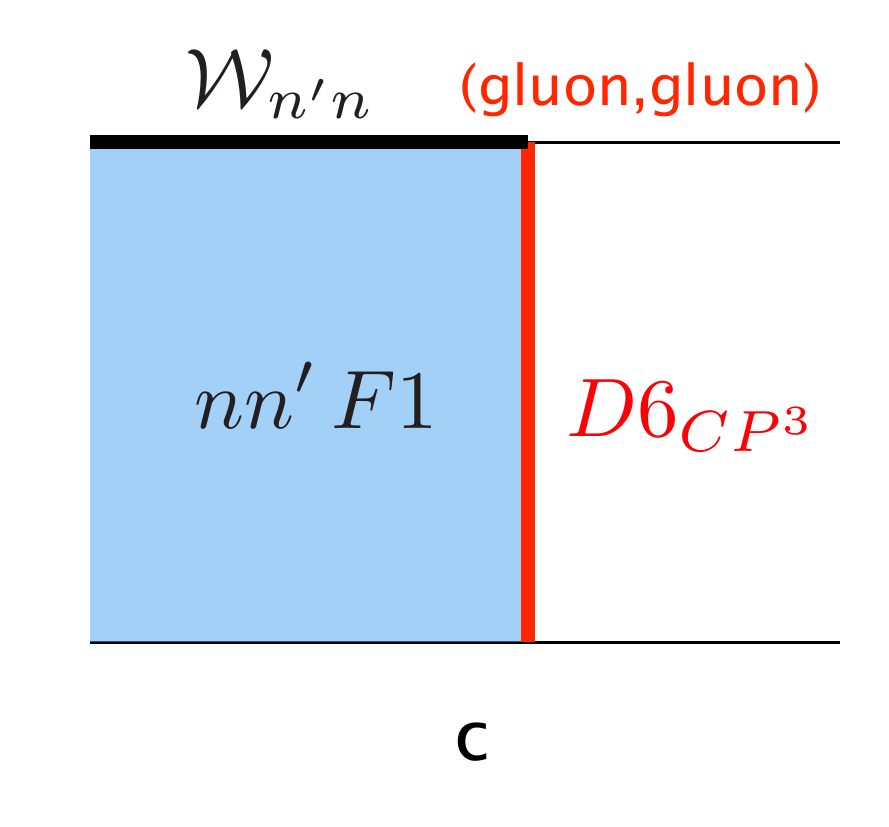} 
\caption{$AdS_4$ dual of $(SU(N)_k\times SU(N)_{-k})/\mathbb{Z}_{n'}$: (a) $n$-dressed-monopole and di-baryon operators. (b) Chiral ring relation. (c) $N=nn'$ fundamental Wilson lines screened by gluons.}
\label{AdS4-5}
\end{figure}

The Dirichlet boundary condition for $A_{D4}$ allows wrapped D4-branes to end on the boundary, 
giving the di-baryon operators ${\cal B}^\ell$.
The $\mathbb{Z}_n$ restricted-free, or equivalently the $\mathbb{Z}_{n'}$ fixed, boundary condition for $A_{D0}$,
allows also D0-branes to end on the boundary in multiples of $n$. 
These correspond to the dressed monopoles ${\cal M}_{n\ell}$.
The chiral ring relation is again described in the bulk as a fully wrapped Euclidean NS5-brane.
The $\mathbb{Z}_n$ fixed boundary condition for $B$ allows string worldsheets to end on the boundary in multiples of $n'$,
describing the Wilson lines ${\cal W}_{n'\ell}$,
with $n$ of these multiples screened by a wrapped D6-brane.
See Fig.~\ref{AdS4-5} for illustrations.
All of this agrees with the properties of $(SU(N)_k\times SU(N)_{-k})/\mathbb{Z}_{n'}$ theory shown in Table~\ref{FieldTheoryProperties}.

\section{Conclusions and outlook}

We have shown that Type IIA string theory in $AdS_4\times \mathbb{C}P^3$,
or equivalently M-theory in $AdS_4\times S^7/\mathbb{Z}_k$, incorporates a larger class
of three-dimensional ${\cal N}=6$ superconformal field theories than was previously appreciated.
As in the case of 4d ${\cal N}=4$ SYM theories, the different 3d ${\cal N}=6$ theories correspond to different boundary 
conditions at the boundary of $AdS_4$ imposed on the bulk gauge fields, and the allowed boundary conditions are constrained by a specific topological 
term in the bulk supergravity theory.
The resulting holographic dualities generalizing the case of the $U(N)_k\times U(N)_{-k}$ theory were shown in Table~\ref{Dualities} above.

If one were to single out one ${\cal N}=6$ theory as the ``mother" theory, the analog of the $SU(N)$ theory in four dimensions, it would 
be the $(SU(N)_k\times SU(N)_{-k})/\bZ_{N}$ theory, which is equivalently formulated as the $(U(N)_k\times U(N)_{-k} )/\bZ_k$ theory.
This theory is dual to the $AdS_4$ background with the ``standard" Dirichlet boundary condition for both one-form gauge fields.
It enjoys an ordinary global symmetry given by $G_0=U(1)\times \mathbb{Z}_{\gcd(N,k)}$, and has two types of charged local operators 
carrying $U(1)$ charges in the ratio $k/N$, one of which is also charged under the discrete $\mathbb{Z}_{\gcd(N,k)}$.
All other ${\cal N}=6$ SCS theories with equal ranks of the two gauge groups,
including the original $U(N)_k\times U(N)_{-k}$ theory, are obtained by gauging a discrete subgroup of $G_0$.
This procedure has two effects in general.
It removes from the spectrum the subset of local operators that are charged under the discrete subgroup,
and at the same time introduces line operators that are allowed by Dirac quantization.
The line operators are charged under a one-form symmetry given by the same discrete subgroup that was gauged.

There are a number of interesting directions for further exploration.
First, it is not clear that we have exhausted all the allowed boundary conditions, in the $AdS_4$ background that we discussed, 
that preserve ${\cal N}=6$ supersymmetry.
A more careful analysis of the boundary conditions consistent with the low energy bulk theory and with ${\cal N}=6$ supersymmetry is necessary.
Given the tight constraints imposed on 3d ${\cal N}=6$ Chern-Simons theories, it would be surprising to find new ${\cal N}=6$ boundary conditions.
But of course those may correspond to 3d SCFT's that do not have a (Chern-Simons) Lagrangian description.

Second, the theories discussed in this paper do not in fact exhaust the list of ${\cal N}=6$ Super-Chern-Simons theories.
There is another class of ${\cal N}=6$ theories with gauge groups $U(N+M)_k\times U(N)_{-k}$ with $1\leq M \leq k-1$ \cite{Aharony:2008gk},
and some discrete quotients thereof \cite{Tachikawa:2019dvq}.\footnote{There are also two special theories with ${\cal N}=6$ supersymmetry 
with gauge groups $SU(N)_k\times U(1)_{-k}$ and $USp(2N)_k\times O(2)_{-2k}$.
However since the rank of the second group is finite in the large $N$ limit we do not expect the supergravity approximation to be valid in these cases.}
The $U(N+M)_{k}\times U(N)_{-k}$ theories were argued to be dual to the $AdS_4\times \mathbb{C}P^3$ background of Type IIA string theory with an additional RR flux and a holonomy for the $B$ field,\footnote{The
$\frac{1}{2}$ shift did not appear in \cite{Aharony:2008gk}, and is there also for $M=0$.
This shift is required in order to cancel an anomalous half-integer tadpole on the D4-brane wrapping $\mathbb{C}P^2$,
that originates from the fact that $\mathbb{C}P^2$ does not admit spin structure but does admit spin$_c$ structure \cite{Aharony:2009fc}.}
\be
\int_{\mathbb{C}P^2} \frac{F_4}{2\pi} = M \,, \;\; 
\int_{\mathbb{C}P^1} \frac{B}{2\pi} = \mbox{} -\frac{M}{k} + \frac{1}{2} \,.
\ee
In the M-theory description this corresponds to a discrete holonomy of the 3-form potential over the torsion 3-cycle in 
$H_3(S^7/\mathbb{Z}_k,\mathbb{Z})=\mathbb{Z}_k$.
It would be interesting to extend the analysis of boundary conditions to this background, especially in view
of the fact that there is no ${\cal N}=6$ version of the $SU(N+M)_k\times SU(N)_{-k}$ theory for $M\neq 0$, and 
in view of the additional constraints that are imposed 
on the allowed discrete quotients of the $U(N+M)_k\times U(N)_{-k}$ theory \cite{Tachikawa:2019dvq}.

Finally, there are many more examples of $AdS_4/CFT_3$ pairs with less supersymmetry in which one can 
study the role of boundary conditions. For example with ${\cal N}=5$ supersymmetry 
we have the orientifold theories $USp(2N+2M)_k\times O(N)_{-2k}$, which have a relatively simple bulk dual \cite{Aharony:2008gk}.
It would be interesting to work out the ${\cal N}=5$ version of the story.

\medskip

\noindent{\bf Acknowledgments}
OB would like to thank a number of people with which he has had illuminating conversations about some of the issues presented in this paper:
Ofer Aharony, Shinji Hirano, Po Shen Hsin, Igor Klebanov, Neil Lambert, and Silviu Pufu.
OB would also like to thank the Aspen Center for Physics and to specifically acknowledge the Winter 2019 program on 
``Higher Symmetries: Theory and Applications"
where some of the ideas presented here began to solidify.
OB is supported in part by the Israel Science Foundation under grant No. 1390/17.
YT is supported by in part supported  by WPI Initiative, MEXT, Japan at IPMU, the University of Tokyo,
and in part by JSPS KAKENHI Grant-in-Aid (Wakate-A), No.17H04837 
and JSPS KAKENHI Grant-in-Aid (Kiban-S), No.16H06335. GZ is supported in part by World Premier International Research Center Initiative (WPI), MEXT, Japan, by the ERC-STG grant 637844-HBQFTNCER and by the INFN.

\appendix
\section{Monopole operators}\label{app:monopole}

A monopole operator in three dimensions is the reduction of a four-dimensional 't Hooft line operator on a circle.
It is defined by the magnetic flux on the two-sphere surrounding it.
In general the spectrum of monopoles and the spectrum of Wilson lines (or equivalently allowed charges) is constrained
by Dirac quantization, and depends on the precise global structure of the gauge symmetry.
Roughly speaking a smaller gauge group restricts the Wilson line spectrum more, and therefore restricts the monopole spectrum less.
Here are some examples.

For a $U(1)$ gauge field, Dirac quantization requires 
\be
\label{U(1)monopole}
H = \frac{1}{2\pi} \int_{{S^2}} F = m \in \mathbb{Z} \,.
\ee
The integer $m$ is a conserved charge corresponding to a topological $U(1)$ symmetry with conserved current $j=*F$.
In the presence of a CS term
\be
{\cal L}_{CS} = \frac{k}{4\pi} A\wedge dA
\ee
an $m$-monopole operator carries an electric charge $q=km$.

For an $SU(N)$ gauge field, a monopole is defined by the the magnetic fluxes in the Cartan subgroup,
\be
\label{SU(N)monopole}
H = \mbox{diag}(m_1,m_2,\ldots,m_N)
\ee
with $\sum_{i=1}^N m_i = 0$. Using the Weyl symmetry we can order the fluxes as $m_1\geq m_2 \geq \cdots \geq m_N$ 
without a loss of generality. Since Wilson lines are permitted in all representations of $SU(N)$, Dirac quantization requires $m_i\in\mathbb{Z}$.
These monopoles do not carry a conserved charge. The current $j=*F$ is not gauge invariant.
In the presence of a CS term
\be
\label{SU(N)CS}
{\cal L}_{CS} = \frac{k}{4\pi} \mbox{Tr}(AdA - \frac{2i}{3}A^3)
\ee
the above monopole transforms in an $SU(N)$ representation given by a Young diagram with $N-1$ rows, where the $i$th 
row has $k(m_i-m_N)$ boxes.

\begin{figure}[h!]
\center
\includegraphics[width=0.4\textwidth]{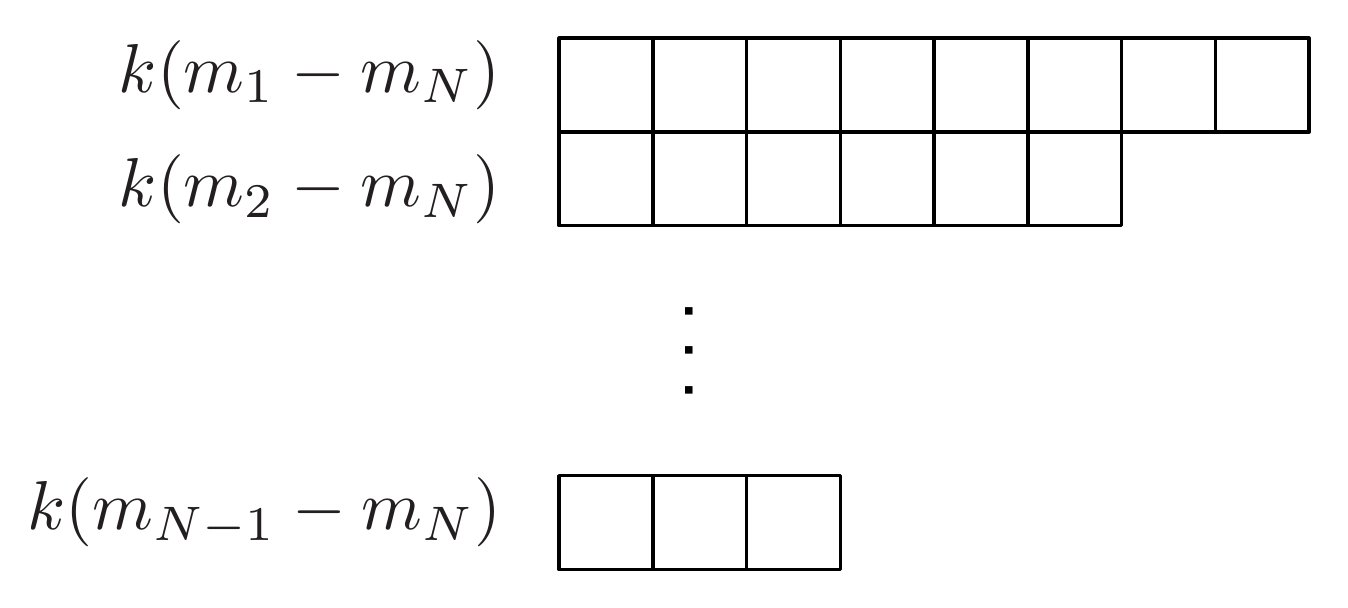} 
\caption{$SU(N)$ representation of an $SU(N)$ monopole in the presence of a level $k$ CS term.}
\label{YoungTableaux}
\end{figure}

For $SU(N)/\bZ_N$ there are additional possibilities for monopoles, since Wilson lines are permitted only in representations 
that are invariant under the center of $SU(N)$, namely in $N$-fold products of the fundamental representation.
This is equivalent to gauging the electric $\mathbb{Z}_N$ global one-form symmetry of the $SU(N)$ theory.
In addition to (\ref{SU(N)monopole}) one also has monopoles given by the magnetic fluxes:
\be
\label{SU(N)/ZNmonopoles}
H = m\, \mbox{diag}\left(\frac{N-1}{N},-\frac{1}{N},\dots,-\frac{1}{N}\right)\,.
\ee
This generates a gauge transformation given by $e^{i2\pi H}$, which is an element of the center $\bZ_N$, and is therefore the identity in $SU(N)/\mathbb{Z}_N$.
There is no gauge-invariant conserved current, but the monopoles carry a conserved charge taking values in $\mathbb{Z}_N$.
This is the reduction of the magnetic one-form $Z_N$ symmetry of the 4d theory, and can be seen
for example from $\pi_1(SU(N)/\mathbb{Z}_N) = \mathbb{Z}_N$.
In the presence of a level $k$ CS term (\ref{SU(N)CS}) the new monopoles transform simply as the symmetric $mk$-fold product
of fundamentals $({\bf N}^{mk})_{sym}$.

Finally consider the case of $U(N)$. 
As a simple generalization of the $U(1)$ case, the $U(N)$ monopoles take the form
\be
\label{U(N)monopole}
H_{U(N)} = (m_1,m_2,\ldots,m_N)
\ee
where $m_i\in \mathbb{Z}$ and $m_1\geq m_2 \geq \cdots \geq m_N$, without a condition on the sum.
These carry a conserved charge given by $\sum m_i$ corresponding to the conserved current $j = *\mbox{Tr}(F)$.
It is also useful to describe these in terms of $SU(N)$ monopoles.
Locally $U(N)$ is similar to $SU(N)\times U(1)$, but the precise global relation is $U(N)=(SU(N)\times U(1))/\mathbb{Z}_N$,
where $\mathbb{Z}_N$ acts simultaneously as the center of $SU(N)$ and $e^{-2\pi i k/N}\in U(1)$.
In particular a Wilson line in the ${\bf N}$ of $SU(N)$ must also have a unit of charge under $U(1)$.
From this point of view one can have $U(1)$ monopoles (\ref{U(1)monopole}),
$SU(N)$ monopoles (\ref{SU(N)monopole}), , and also monopoles of the form
\be
\label{U(N)monopole2}
H_{SU(N)} &=& m\, \mbox{diag}\left(\frac{N-1}{N},-\frac{1}{N},\dots,-\frac{1}{N}\right)\nonumber\\
H_{U(1)} &=& \mbox{} \frac{m}{N}
\ee
where $m\in \mathbb{Z}$.
This is easily seen to be an equivalent description to (\ref{U(N)monopole}) by decomposing the $U(N)$ gauge field into an $SU(N)$ gauge field and
a $U(1)$ gauge field as ${\cal A} = A + a {\bf I}$.
In particular the monopole above takes the form $H_{U(N)} = (m,0,\ldots,0)$.
In the presence of CS terms the monopoles again acquire gauge charges.
In general the $SU(N)$ and $U(1)$ CS levels may be different, but the difference must be a multiple of $N$.
The CS action for the so-called $U(N)_{k,k+Nk'}$ theory is given by 
\be
{\cal L}_{CS_{k,k+Nk'}} &=& \frac{k}{4\pi} \mbox{Tr}({\cal A}d{\cal A} - \frac{2i}{3}{\cal A}^3) 
+ \frac{k'}{4\pi} \mbox{Tr}({\cal A}) d \mbox{Tr}({\cal A}) \nonumber \\
&=& 
\frac{k}{4\pi} \mbox{Tr}({A}d{A} - \frac{2i}{3}{A}^3) + \frac{N(k+Nk')}{4\pi} ada \,.
\ee
For $k'=0$ the CS levels are the same and this describes the $U(N)_{k,k}$ theory.
The general $U(N)$ monopole (\ref{U(N)monopole}) transforms in the $SU(N)$ representation shown in Fig.~\ref{YoungTableaux},
and carries a $U(1)$ charge $q = (k+Nk')\sum_{i=1}^N m_i$.
In particular the monopole in (\ref{U(N)monopole2}) transforms in the $({\bf N}^{mk})_{sym}$ and has $q=(k+Nk')m$.

\bibliographystyle{ytphys}
\bibliography{ref}
\end{document}